\documentclass[%
 reprint,
superscriptaddress,
 amsmath,amssymb,
 aps,
prb,
]{revtex4-2}

\usepackage{graphicx}
\usepackage{dcolumn}
\usepackage{bm}
\usepackage[caption=false]{subfig}

\usepackage[dvipsnames]{xcolor}
\usepackage{tzplot} 
\usepackage{tikz}
\usepackage{tikz-3dplot} 
\usepackage{rotating}
\usetikzlibrary{3d}
\usepackage{url} 
\usepackage{bm}

\definecolor{darkred}{rgb}{0.5,0,0}
\definecolor{darkblue}{rgb}{0,0,0.5}
\definecolor{firebrick}{rgb}{0.75,0.125,0.125}  
\definecolor{cegla}{rgb}{0.98, 0.33, 0.25}
\definecolor{darkgreen}{RGB}{6, 23, 14}
\definecolor{niebo}{rgb}{0,0.9,1}
\definecolor{royalblue}{rgb}{0  0.5  15}
\definecolor{trawa}{rgb}{0.1  0.3  0.1}
\definecolor{zielnieb}{rgb}{0.18, 0.5, 0.5}
\definecolor{turkus}{rgb}{0.5, 0.4, 0.5}
\definecolor{kasztan}{RGB}{100,0,0}
\definecolor{midnightblue}{RGB}{11, 11, 49}
\definecolor{grafit}{RGB}{120,120,120}
\definecolor{sliwa}{RGB}{68, 46, 60}
\usepackage{hyperref}
\hypersetup{
	pdftitle = {}, 
	pdfauthor = {}, 
	colorlinks,
	linktoc=page,
	linkcolor=grafit,
	citecolor=grafit,
	filecolor=grafit,
	urlcolor=grafit
}

\newcommand{\Tstrut}{\rule{0pt}{2.7ex}}
\newcommand{\Bstrut}{\rule[-1.2ex]{0pt}{0pt}}

\begin{document}

\preprint{APS/123-QED}

\title{From many valleys to many topological phases --- quantum anomalous {Hall} effect in {IV--VI} semiconductor quantum wells}

\author{Szymon Majewski}
\email{majewski@MagTop.ifpan.edu.pl}
\affiliation{%
 International Research Centre MagTop, Institute of Physics,
Polish Academy of Sciences, Aleja Lotnikow 32/46, PL-02668 Warsaw, Poland
}%
\author{Michał Wierzbicki}
\email{michal.wierzbicki@pw.edu.pl}
\affiliation{
 Warsaw University of Technology, Faculty of Physics, Koszykowa 75, PL-00-662 Warsaw, Poland
}%
\author{Tomasz Dietl}%
\email{dietl@MagTop.ifpan.edu.pl}
\affiliation{%
 International Research Centre MagTop, Institute of Physics,
Polish Academy of Sciences, Aleja Lotnikow 32/46, PL-02668 Warsaw, Poland
}%

\date{\today}         

\begin{abstract}
Consistent with prior qualitative expectations for group IV--VI topological crystalline insulators, this work demonstrates, based on band structure and Chern number calculations, that  Pb$_{1-x}$Sn$_x$Se/(PbSe)$_{1-y}$(EuS)$_y$ quantum wells constitute a promising and viable platform for realizing a variety of quantum anomalous Hall phases. The proposed basis transformation procedure for the multiband $\bm{\mathit{k}} \cdot \bm{\mathit{p}}$ Hamiltonian enables the treatment of wells grown along arbitrary crystallographic directions while explicitly accounting for the anisotropy of the material’s isoenergetic surfaces. Numerical studies of $\langle 111\rangle$-, $\langle 110\rangle$- and $\langle 001\rangle$-oriented quantum wells predict attainable Chern numbers with magnitudes ranging from $1$ to $4$, depending on the quantum well width,  Sn content, and relative orientation of the four projected $\mathrm{L}$ valleys with respect to the growth direction. The results further indicate that appropriate strain compensation is required to achieve high-quality quantization of the Hall conductance.        
\end{abstract}

\maketitle


\section{Introduction}
\label{sec:intro}

Phenomena belonging to the quantum Hall effects family play a significant role in modern condensed-matter physics, serving as the source of many new theoretical concepts, and driving the development of fields such as metrology, spintronics, energy efficient electronics, and quantum computation \cite{Klitzing:2020_NRP}. Moreover, the recent discovery of a strong influence of the vacuum fluctuations on the integer and fractional quantum Hall effects in a photon cavity \cite{Enkner:2025_N} opens the door for searches of coupling to other quantum fluctuations in various materials and experimental settings. In those contexts, the quantum anomalous Hall effect (QAHE) \cite{Liu:2008_PRL, Chang:2013_Science, Chang:2023_Rev_Mod_Phys} occupies a prominent place: the quantization of the Hall conductivity, given by $\sigma_{\text{H}} = \mathcal{C}e^2/h$ (with $\mathcal{C}$ an integer), results from the nontrivial topology of the band structures of 2D crystals with broken time-reversal symmetry (TRS), rather than from the formation of Landau levels as in the ``conventional" quantum Hall effect. Of particular interest is the prospect of a quantum electrical resistance standard functioning at zero magnetic field which, in tandem with the Josephson-effect quantum voltage standard, enables a quantum standard for the ampere \cite{Picard:2016_PRX,Rodenbach:2025_NE}. To date, thin films of topological insulator $($Bi,Sb$)_{2}$Te$_3$ doped with Cr or V \cite{Chang:2013_Science,Rodenbach:2025_NE, Chang:2015_Nat_Mater, Okazaki:2022_NP, Molenkamp:2024_Nat_Electron} or sandwiched between (Zn,Cr)Te barriers \cite{Watanabe:2019_APL}, along with multilayer graphene \cite{ZLu:2024_Nat}, are examples of the systems in which the QAHE has been observed. Typically, coercivity is those samples is below $1\,\mathrm{T}$, and a standard permanent magnet can serve to align magnetic domains \cite{Okazaki:2022_NP}. There exist also theoretical works considering theoretically a possibility of the QAHE in quantum wells of HgTe doped with Mn \cite{Liu:2008_PRL} or with Cr or V  \cite{Sliwa:2024_PRB,Cuono:2024_PRB} awaiting for the experimental verification. Interestingly, the uppermost hole Landau level at the topological phase transition in HgTe-based quantum wells, gives rise to a broad quantized plateau appearing already in fields achievable by permanent magnets \cite{Shamim:2020_SA, Yahniuk:2021_arXiv}. Importantly, Hall devices based on $($Bi,Sb,Cr$)_{2}$Te$_3$ and $($Bi,Sb,V$)_{2}$Te$_3$ show already the quantization accuracy relevant for metrology \cite{Rodenbach:2025_NE, Okazaki:2022_NP, Molenkamp:2024_Nat_Electron}, but only at extremely low currents and temperatures below $50\,\mathrm{mK}$, owing to the high defect concentration giving rise to parallel hopping conduction at non-zero source-drain electric fields and temperatures \cite{Okazaki:2022_NP}.

\begin{figure*}[t]
	\centering
	\subfloat[\label{subfig:IV-VIs_a}]{\includegraphics[scale=1.0]{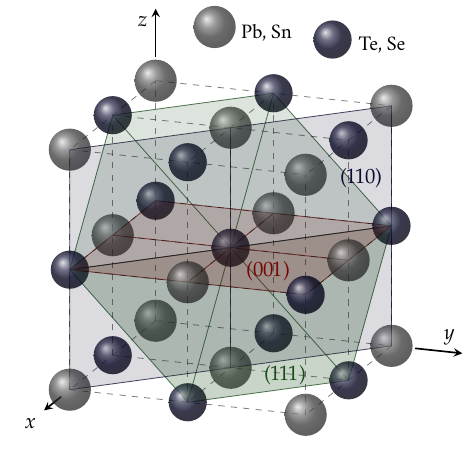}}\quad\quad
    \subfloat[\label{subfig:IV-VIs_b}]{\includegraphics[scale=1.0]{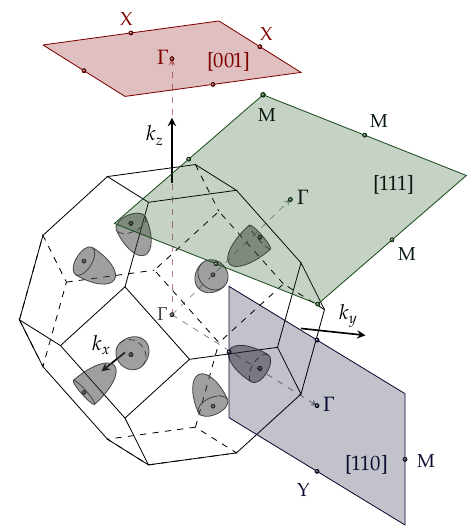}}
 \caption{(a) Rock-salt crystal structure of IV--VI compounds with representative  $\{001\}$, $\{110\}$ and $\{111\}$ planes; the symmetries of these planes ensure topological protection of gapless surface states \cite{Hsieh:2012_Nat_Commun}. (b) Projections of the three-dimensional FBZ of IV--VI semiconductors onto the quantum well growth directions considered in this work. Schematic constant-energy surfaces near the L-point band edges --- approximately spheroidal \cite{Nimtz:1983} --- are also shown; their anisotropy and reciprocal-space orientations induce lifting of the L-valley degeneracy upon FBZ projection onto $\langle 111 \rangle$ and $\langle 110 \rangle$ directions.} 
\label{fig:IV-VIs}
\end{figure*} 

In this paper, we present quantitative theoretical predictions on the QAHE in quantum wells (QWs) of IV--VI semiconductors that crystallize in the rock-salt cubic structure (cf. Fig.~\ref{subfig:IV-VIs_a}). The mineral galena, i.e., PbS, was arguably the first functional semiconductor ever used \cite{Emerson:1997_IEEE,Russer:2009_EuMC}, but IV--VI compounds continue to attract a lot of attention as efficient thermoelectric materials \cite{Gao:2025_S,Ginting:2025_SSS} and infrared detectors \cite{Gluch:2025_APL}. Importantly for this work, (Pb,Sn)(Se,Te) for Sn concentrations above a certain threshold, constitute members of the class of topological crystalline insulators (TCIs) \cite{Hsieh:2012_Nat_Commun,Dziawa:2012_Nat_Mater,Tanaka:2012_NP,Xu:2012_NC,Zgang:2020_PRM}, for which the presence of an interfacial gapless Dirac cone and a parity anomaly was already discussed theoretically in the 1980s \cite{Volkov:1985_JETPL,Fradkin:1986_PRL,Korenman:1987_PRB}. The band inversion occurs in Sn-rich compounds, as the relativistic downward shift of $s$ orbitals is smaller in Sn compared to heavier Pb \cite{Ye:2015_npjCM}. As a result, $s$-$p$ hybridization shifts, at the $\mathrm{L}$ point of the Brillouin zone, bands of anion $p$ orbitals {\em above} bands of cation $p$ orbitals for a sufficiently large Sn content \cite{Dziawa:2012_Nat_Mater,Ye:2015_npjCM}. Furthermore, as already noted within a model Hamiltonian approach \cite{Fang:2014_PRL}, the location of the energy gap at the four $\mathrm{L}$ points of the three-dimensional first Brillouin zone (FBZ) permits the realization of a system manifesting distinct QAH phases in (Pb,Sn)(Se,Te) doped with Mn or Cr --- namely, phases characterized by different Chern numbers  $|\mathcal{C}|\in\{ 1,\dots,4 \}$. The QAHE with  $|\mathcal{C}| = 2$ was also predicted for a monolayer of PbTe or SnTe sandwiched between NaCl and subject to a ferromagnetic proximity effect from magnetic adatoms or a substrate \cite{Niu:2015_PRB}.

Our search for the QAH phases in (PbSe)$_{1-y}$(EuS)$_y$\,| Pb$_{1-x}$Sn$_x$Se\,|\,(PbSe)$_{1-y}$(EuS)$_y$ QWs within multiband envelope function formalism is motivated by and rely on four developments:

First, several groups have already mastered the epitaxial growth of (Pb,Sn)Se quantum wells with Sn content approaching the topological region and carrier densities low enough for shifting the Fermi level to the gap by gating and/or illumination \cite{Wang:2020_PRB,Krizman:2021_PRB,Kazakov:2025_PRB,Wang:2015_AM}. These studies have also reconfirmed the crucial role of epitaxial strain, whose compensation is necessary to maintain a non-zero bandgap in those multivalley systems. Carrier confinement, strain compensation, and structure compatibility can be achieved by barriers containing wide-gap rock-salt EuS. 

Second, EuS not only has a crystal structure compatible with IV--VI TCIs, but also exhibits ferromagnetic ordering of Eu spins $S = 7/2$ below $16.6\,\mathrm{K}$ \cite{Bohn:1981_JAP}. However, for strain compensation reasons, instead of EuS, we consider (Pb,Eu)(S,Se) as a TRS-breaking medium. In that magnetic material, the dilution and competing ferromagnetic and antiferromagnetic interactions can result in a spin-glass phase, as in (Sr,Eu)S \cite{Maletta:1982_JAP}.  Nevertheless, since the magnitude of antiferromagnetic Eu-Eu exchange integrals attains only $0.24\,\mathrm{K}$ in dilute (Pb,Eu)Se \cite{Bindilatti:1998_PRB,Gorska:2022_pssb}, and is presumably smaller or even  ferromagnetic in (Pb,Eu)(Se,S), we expect a full saturation of Eu spin magnetization in fields achievable by permanent magnets. The presence of Eu only in the barriers is beneficial, as the QAHE is thought to result from the effect of TRS breaking on the topological interfacial states. 

Third, comprehensive magnetooptical studies \cite{Bauer:1992_Sem_Scien_Tech} and theoretical modeling \cite{Bauer:1992_Sem_Scien_Tech,Dietl:1994_PRB} of Mn- and Eu-doped PbSe and PbTe epitaxial layers have provided quantitative information on $\bm{\mathit{k}} \cdot \bm{\mathit{p}}$ parameters and exchange coupling between band carriers and localized spins. This information allows us for experimentally relevant predictions on the QAHE. In particular, the appropriate magnitudes and signs of the relevant exchange integrals are essential for the band inversion in one spin channel, the key condition for the appearance of the QAHE \cite{Liu:2008_PRL}. 

Finally, compared to bismuth-antimony chalcogenides, lead-tin chalcogenides are characterized by even higher magnitude of the dielectric constant and typically lower defect concentrations. Furthermore, the multivalley band structure suggests a possibility of achieving various Chern numbers, up to $|\mathcal{C}| = 4$. All those properties may be pertinent to shifting the accurate Hall conductivity quantization to higher currents and temperatures.

Our results reveal that for the adopted values of the band structure parameters and the experimentally feasible quantum well architectures, the QAHE can be observed in IV--VI TCIs in magnetic fields of standard permanent magnets. The presented topological phase diagrams for Pb$_{1-x}$Sn$_x$Se/(PbSe)$_{1-y}$(EuS)$_y$ QWs demonstrate  that the quantized resistance values from $h/e^2$ to $h/4e^2$ are achievable by selecting appropriate crystallographic orientations of the growth directions $(\langle 111\rangle$, $\langle 110\rangle$, $\langle 001\rangle)$, quantum well thicknesses,  Sn concentrations, and strain compensation. In particular, strain compensation prevent the gap closure for $\langle 111\rangle$ and  $\langle 110\rangle$ QWs. Moreover, tensile strain, i.e., using EuSe instead of EuS, would shift the topological phase transition to higher Sn content $x$ at which the cubic phase may no longer persist. Our theoretical predictions call for experimental verification to assess how the thermal and current stability of quantization accuracy compares with other material families and whether interfacing with superconductors is feasible and leads to new functionalities.

\section{Numerical modelling}
\label{sec:modelling}

\begin{figure}[t]
	\centering
	\includegraphics[scale=0.95]{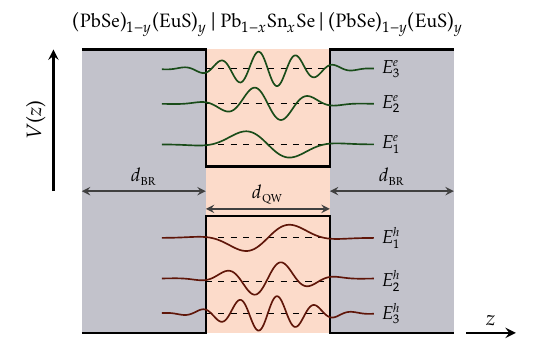}
  \vspace{-0.5cm}
 \caption{Schematic of the modeled quantum wells. The conduction and valence band edges, $E_{c}$ and $E_{v}$, are sketched along the $z$-axis, and the electron and hole states, $E^{e}_n$ and $E^{h}_n$, are indicated schematically. The energy gap in this system is $E_{\text{g}} = E^{e}_1 - E^{h}_1$.} 
\label{fig:QW}
\vspace{-0.35cm}
\end{figure}

\subsection{Band structure calculations}
\label{subsec:bandcalc}

We determine the band structure and topological invariants of IV--VI quantum wells with magnetic barriers for the $\langle 111\rangle$, $\langle 110\rangle$, and $\langle 001\rangle$ crystallographic orientations. To this end, we adopt the four-band $\bm{\mathit{k}}\cdot\bm{\mathit{p}}$ model recently developed in the context of experimental studies of (Pb,Sn)Se QWs grown along the $\langle 111\rangle$ crystal axis \cite{Kazakov:2025_PRB}, which we generalize to arbitrary growth directions by an appropriate basis transformation.

A convenient point of departure for presenting the $\bm{\mathit{k}}\cdot\bm{\mathit{p}}$ model is the Hamiltonian for the $[111]$ quantum well. In the basis
\begin{equation}
\{ |\lambda_j\rangle \} \equiv  \left\{ |L_6^{+}(\uparrow)\rangle, |L_6^{+}(\downarrow)\rangle, |L_6^{-}(\uparrow)\rangle, |L_6^{-}(\downarrow)\rangle \right\}
\label{eq:basis_phi}
\end{equation} 
(the symbols $L_6^{\pm}$ denote the relevant double-group representations; cf. e.g. \cite{Dimmock:1964_PhysRev}) for the in-plane wave vector $\bm{\mathit{k}}=[k_x,k_y]^{\mathrm{T}}$ (with coordinate axes $\left\{ [\bar{1}\bar{1}2], [1\bar{1}0], [111]\right\}$), it comprises the following terms:
\begin{equation}
\hat{\mathcal{H}}(\bm{\mathit{k}}) = \hat{\mathcal{V}}(z) + \hat{\mathcal{H}}_{\text{SO}}(\bm{\mathit{k}}) + \hat{\mathcal{H}}_{\text{Z}} + \hat{\mathcal{H}}_{\text{exc}}\, .
\label{eq:kp_Ham}
\end{equation} 
The first term, $\hat{\mathcal{V}}(z) = \mathrm{diag}(V_v(z), V_v(z),V_c(z),V_c(z))$, describes the valence- and conduction-band-edge profiles $V_v(z)$ and $V_c(z)$ along the unit vector $\hat{\bm{\mathit{e}}}_z$~$\parallel$~$[111]$ defining the growth direction of the quantum well (cf. Fig.~\ref{fig:QW}) and incorporates strain effects arising from lattice mismatch between the layers. A detailed account of the strain treatment and explicit expressions for $V_{c,v}(z)$ is provided in Appendix~\ref{sec:appendix_A}. 

The second contribution, representing the effective coupling to proximate bands in the presence of spin-orbit interaction (SOI), is given by
\begin{equation}
\hat{\mathcal{H}}_{\text{SO}}(\bm{\mathit{k}}) = \left[ \begin{array}{cc}
\hat{H}^{v}_{\text{SO}}(\bm{\mathit{k}}) & \hat{H}^{vc}_{\text{SO}}(\bm{\mathit{k}}) \\
\hat{H}^{cv}_{\text{SO}}(\bm{\mathit{k}}) & \hat{H}^{c}_{\text{SO}}(\bm{\mathit{k}})
\end{array} \right] ,
\end{equation}
with the block-diagonal entries
\begin{equation}
\hat{H}^{v,c}_{\text{SO}}(\bm{\mathit{k}}) = \mp \left( \hbar^2\frac{k_x^2 + k_y^2}{2m_{t}^{v,c}} + \frac{\hbar^2\hat{k}_z^2}{2m_{l}^{v,c}} \right)\hat{\tau}_0\, ,
\end{equation}
and the off-diagonal blocks
\begin{equation}
(\hat{H}^{cv}_{\text{SO}})^{\dagger}(\bm{\mathit{k}}) = \hat{H}^{vc}_{\text{SO}}(\bm{\mathit{k}}) = \hbar v_t (k_x \hat{\tau}_x + k_y \hat{\tau}_y) + \hbar v_l \hat{k}_z \hat{\tau}_z\, .
\end{equation}
In these expressions $\hat{\tau}_j$, $j\in\{ x,y,z \}$, denote the Pauli matrices and $\hat{\tau}_0$ -- the $2\times 2$ identity matrix. The parameters $m_{t,l}^{v,c}$ correspond to the transverse $(t)$ and longitudinal $(l)$ relative to the $[111]$ direction contributions to effective masses in the valence $(v)$ and conduction $(c)$ bands brought about by remote bands. The velocities $v_{t,l}$ parametrize the $\bm{\mathit{k}}\cdot\bm{\mathit{p}}$ perturbation including SOI; they replace the conventional momentum-matrix elements according to $P_{t,l} = m_0v_{t,l}$, where $m_0$ is the free-electron mass. To account for electron confinement along the $[111]$ direction, the $z$-component of the wave vector in the above expressions is replaced by the operator $\hat{k}_z$, which in a magnetic field $\bm{\mathit{B}}$ also depends on the vector potential $\bm{\mathit{A}} = (\bm{\mathit{B}}\times\bm{\mathit{r}})\ /\ 2$ with $\bm{\mathit{r}} = [ x, y, z ]^{\mathrm{T}}$; explicitly,
\begin{equation}
\hat{k}_z \equiv ( -\mathrm{i}\hbar\partial_z + eA_z )\ /\ \hbar\, .
\label{eq:kz}
\end{equation}

\begin{table*}[t]
\caption{\label{tab:kp_params} Values of the $\bm{\mathit{k}}$ $\cdot$ $\bm{\mathit{p}}$ model parameters employed in the calculations \cite{Kazakov:2025_PRB, Nimtz:1983, Kritzman:2018_PRB, Dietl:1994_PRB}. The quantities $m_{l,t}^{v,c}$ are expressed in units of the free-electron mass $m_0$. Due to the scarcity of experimental data, identical parameter values were adopted for the well and barrier materials (except for the exchange constants $\mathcal{J}_{v,c}$, which are nonzero only within the barriers).}
\begin{ruledtabular}
\begin{tabular}{cccccccccccc}
		 $m_{l}^{v}$ & $m_{t}^{v}$ & $m_{l}^{c}$ & $m_{t}^{c}$ & $v_{l}[\mathrm{m}/\mathrm{s}]$ & $v_{t}[\mathrm{m}/\mathrm{s}]$ & $g_{l}^{v}$ & $g_{t}^{v}$ & $g_{l}^{c}$ & $g_{t}^{c}$ & $\mathcal{J}_v[\mathrm{meV}]$ & $\mathcal{J}_c[\mathrm{meV}]$\Bstrut \\  \hline 
		\Tstrut $0.3$ & $0.1$ & $0.3$ & $0.2$ & $4.0\cdot 10^5$ & $4.2\cdot 10^5$ & $-3.3$ & $-0.8$ & $-5.1$ & $-3.5$ & $80$ & $20$ \\ 
	\end{tabular}
\end{ruledtabular}
\end{table*}

The subsequent term represents the Zeeman coupling of the carriers to an external magnetic field $\bm{\mathit{B}}$ $=$ $[B_x,B_y,B_z]^{\mathrm{T}}$ $=$ $B[b_x, b_y, b_z]^{\mathrm{T}}$, leading to an anisotropic (i.e., orientation-dependent) splitting of the spin-resolved bands:
\begin{equation}
\hat{\mathcal{H}}_{\text{Z}} = \left[ \begin{array}{cc}
\hat{H}_{\text{Z}}^{v} & \hat{0} \\
\hat{0} & \hat{H}_{\text{Z}}^{c}
\end{array} \right] ,
\label{eq:HZeeman}
\end{equation}
where $\hat{0}$ denotes the $2\times 2$ zero matrix, and
\begin{equation}
\hat{H}_{\text{Z}}^{v,c} = \mp \frac{\mu_{\text{B}}}{2}\left[ g_t^{v,c}(B_x\hat{\tau}_x + B_y\hat{\tau}_y) + g_l^{v,c}B_z\hat{\tau}_z \right] .
\end{equation}
Here $g_{t,l}^{v,c}$ are the contributions to the effective $g$-factors brought about by remote bands (indexing is analogous to that of $m_{t,l}^{v,c}$). The final contribution to the Hamiltonian (Eq.~(\ref{eq:kp_Ham})) plays a similar, yet substantially more consequential, role in the present model. It implements the exchange coupling of band carriers to magnetic dopants residing in the barrier regions, providing the dominant mechanism for removal of Kramers degeneracy and consequent breaking of TRS. It should be emphasised that the external magnetic field here merely serves to induce magnetization of the barriers. The exchange term may be expressed in a form originating from a Heisenberg-type exchange Hamiltonian describing the interaction between carrier spins and the localized impurity magnetic moments \cite{Bauer:1992_Sem_Scien_Tech}:
\begin{equation}
\hat{\mathcal{H}}_{\text{exc}} = \left[ \begin{array}{cc}
\hat{H}_{\text{exc}}^{v} & \hat{0} \\
\hat{0} & \hat{H}_{\text{exc}}^{c}
\end{array} \right] ,
\label{eq:Hexc}
\end{equation}
where
\begin{equation}
\hat{H}_{\text{exc}}^{v,c} = \frac{1}{2}\Delta_{dsf}^{v,c}\left[ \pm\left( b_x\hat{\tau}_x + b_y\hat{\tau}_y \right) + b_z\hat{\tau}_z \right] .
\label{eq:H_exc_vc}
\end{equation}
The parameter $\Delta^{v,c}_{dsf}$ quantifies the strength of the exchange interaction and scales with the magnetization of magnetic ions in the
barriers, the magnetic-ion concentration $y$ and an exchange constant $\mathcal{J}$, to be understood as an effective \textit{exchange integral}. Assuming paramagnetic behaviour of the Eu ions in the barriers for small $y$, this quantity can be written as  
\begin{equation}
\Delta^{v,c}_{dsf} = y\mathcal{J}_{v,c} S \mathcal{B}_{S}(\xi_B)\, ,\thickspace\thickspace\thickspace \xi_B = \frac{g\mu_{\text{B}}S B}{k_{\text{B}}T}\, ,
\label{eq:Delta_dsf}
\end{equation}
with $g=2$ the $\text{Eu}^{2+}$ Land\'e factor, $S=7/2$ the total spin of the half-filled $4f$ subshell, and $\mathcal{B}_{S}(\xi_B)$ the Brillouin function for the spin $S$. Given the relatively weak anisotropy of the Eu–carrier exchange interaction in PbSe, the model employs a single exchange constant $\mathcal{J}_{v,c}$ for carriers within each band. The values assumed here are $\mathcal{J}_{v} = 80\ \mathrm{meV}$ for holes and $\mathcal{J}_{c} = 20\ \mathrm{meV}$ for electrons \cite{Bauer:1992_Sem_Scien_Tech,Dietl:1994_PRB} (all model parameters are summarized in Table~\ref{tab:kp_params}). Notably, in contrast to certain previous theoretical treatments (e.g. \cite{Kazakov:2025_PRB}), the coupling to conduction-band carriers is retained: it is substantially weaker than the hole coupling but nevertheless comparable in order of magnitude. It is also worth noting that, for $B=0$ and sufficiently large magnetic-ion concentrations $y$, the Brillouin function $\mathcal{B}_{S}(\xi_B)$ entering Eq.~(\ref{eq:Delta_dsf}) may be replaced by a barrier-magnetization coefficient $\eta\in[0,1]$. Crucially, this replacement permits investigation of how the barrier-magnetization orientation --- determined by the components of the unit vector $\bm{\mathit{b}}$ in Eq.~(\ref{eq:H_exc_vc}) --- affects the well band structure, without requiring the vector potential to be included in the definition of $\hat{k}_z$ (Eq.~(\ref{eq:kz})). Inclusion of the vector potential would otherwise be necessary (and technically awkward) for $B\neq 0$ when $\bm{\mathit{B}} \nparallel \hat{\bm{\mathit{e}}}_z$. 

\begin{figure}[b]
	\vspace{-0.3cm}
	\centering
	\includegraphics[scale=0.9]{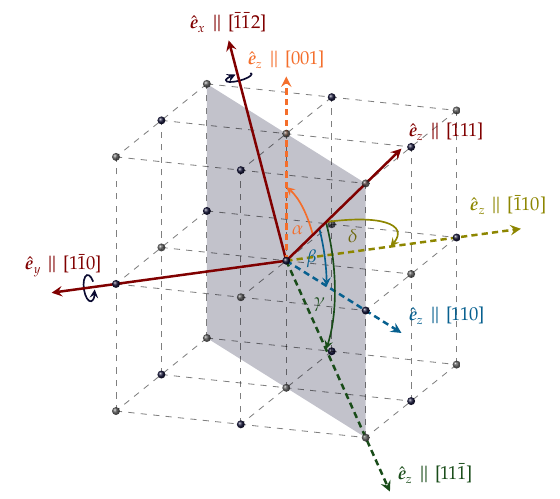}  
  \vspace{-0.2cm}
 \caption{Geometry of the basis (coordinate-system) transformation in the $\bm{\mathit{k}}$~$\cdot$~$\bm{\mathit{p}}$ model. Rotation angles about the respective axes are: \textcolor{Orange}{$\alpha$ $=$ $\arccos(1/\sqrt{3})$ $\approx$ $ 54.7$\textdegree}, \textcolor{MidnightBlue}{$\beta$ $=$ $\arccos(\sqrt{2/3})$ $\approx$ $35.3$\textdegree}, \textcolor{trawa}{$\gamma$ $=$ $\arccos(1/3)$ $\approx$ $70.5$\textdegree} and \textcolor{olive}{$\delta$ $=$ $90$\textdegree}. \textcolor{darkred}{Solid arrows} indicate the coordinate axes directions in which the Hamiltonian (\ref{eq:kp_Ham}) is written. The orientation of the constant-energy ellipsoids with respect to the unit vector $\hat{\bm{\mathit{e}}}_z$ determines the quantization conditions of the energy levels derived from a given valley when electrons are confined along that direction.} 
\label{fig:TR}
\vspace{-0.1cm}
\end{figure}

As illustrated in Fig.~\ref{subfig:IV-VIs_b}, projection of the three-dimensional system onto the $[111]$ axis distinguishes the four $\mathrm{L}$ valleys into a longitudinal $\Gamma$ valley ($[111]$) and three oblique $\mathrm{M}$ valleys ($[\bar{1}11]$,\,$[1\bar{1}1]$,\,$[11\bar{1}]$). The Hamiltonian constructed above accounts for the quantization of energy levels in a quantum well with $\hat{\bm{\mathit{e}}}_z \parallel [111]$ only for the valley whose constant-energy surface is aligned with the coordinate axes, namely the $\Gamma$ valley. In order to analyse the band structure in the vicinity of a particular 
$\mathrm{M}$ point in a manner that properly incorporates the axial anisotropy reflected in the $\bm{\mathit{k}}\cdot\bm{\mathit{p}}$ parameters, one must transform the Hamiltonian (\ref{eq:kp_Ham}) into the basis associated with the appropriate $\mathrm{L}$ point --- i.e., the $\mathrm{L}$ point that projects onto the chosen $\mathrm{M}$ --- relative to the growth direction of the layers. A critical observation is that the foregoing reasoning generalizes to quantum wells grown along other crystallographic directions, notably $[001]$ and $[110]$. Projection of the three-dimensional first Brillouin zone onto the $[001]$ axis maps the four $\mathrm{L}$ valleys  onto two equivalent $\mathrm{X}$ valleys, whereas projection onto $[110]$ maps them onto $\mathrm{Y}$ ($[111]$,\,$[11\bar{1}]$) and $\mathrm{M}$ ($[\bar{1}11]$,\,$[1\bar{1}1]$) valleys. The geometry underlying the discussed procedure is depicted in Fig.~\ref{fig:TR}. 

The transformation consists of a rotation of the spin basis about the axis defined by the unit vector $\bm{\mathit{\epsilon}}$ by an angle $\vartheta$, implemented by the unitary $\mathrm{SU}(2)$ operator  
\begin{equation}
\hat{W}^{\bm{\mathit{\epsilon}}}_{\vartheta} = \exp\left(-\frac{\mathrm{i}\vartheta}{2}\bm{\mathit{\epsilon}}\cdot\hat{\bm{\mathit{\tau}}}\right) \in \mathrm{SU}(2)\, ,
\end{equation}
where $\hat{\bm{\mathit{\tau}}} = [\hat{\tau}_x, \hat{\tau}_y, \hat{\tau}_z]^{\mathrm{T}}$. In light of Fig.~\ref{fig:TR}, $\bm{\mathit{\epsilon}} = \hat{\bm{\mathit{e}}}_y$ for the $\mathrm{M}(111)$ (with $\hat{\bm{\mathit{e}}}_z \parallel [11\bar{1}]$), $\mathrm{X}$ $(\hat{\bm{\mathit{e}}}_z \parallel [001])$ and $\mathrm{Y}$ $(\hat{\bm{\mathit{e}}}_z \parallel [110])$ valleys, whereas $\bm{\mathit{\epsilon}} = \hat{\bm{\mathit{e}}}_x$ for the  $\mathrm{M}(110)$ $(\hat{\bm{\mathit{e}}}_z \parallel [\bar{1}10])$ valley. Rotation angles $\vartheta$ for the individual operations are reported in the caption of Fig.~\ref{fig:TR}. The Hamiltonian in the rotated frame is given by
\begin{equation}
\hat{\mathcal{H}}^{\bm{\mathit{\epsilon}}}_{\vartheta}(\bm{\mathit{k^{\prime}}}) = \hat{T}^{\bm{\mathit{\epsilon}}}_{\vartheta}\hat{\mathcal{H}}(\hat{R}^{\bm{\mathit{\epsilon}}}_{\vartheta}\bm{\mathit{k^{\prime}}})\hat{T}^{\bm{\mathit{\epsilon}}}_{\vartheta}{}^{\dagger}\, ,
\end{equation}
where $\hat{R}^{\bm{\mathit{\epsilon}}}_{\vartheta} \in \mathrm{SO}(3)$ denotes the corresponding rotation in real space, $\bm{\mathit{k^{\prime}}}$ is the wave vector expressed in the new coordinate frame, and the block-diagonal spin rotation operator reads
\begin{equation}
\hat{T}^{\bm{\mathit{\epsilon}}}_{\vartheta} = \left[ \begin{array}{cc}
\hat{W}^{\bm{\mathit{\epsilon}}}_{\vartheta} & \hat{0} \\
\hat{0} & \hat{W}^{\bm{\mathit{\epsilon}}}_{\vartheta}
\end{array} \right] .
\end{equation}
In summary, the employed procedure concerns the Hamiltonians for carriers in a magnetic field, residing in the valleys oblique to the growth direction. A suitable form of these Hamiltonians is obtained by transforming the Hamiltonian (\ref{eq:kp_Ham}) in the coordinate system with the $\hat{\bm{\mathit{e}}}_z$ direction along the long axis of the constant-energy ellipsoid to the form with the $\hat{\bm{\mathit{e}}}_z$ axis along the growth direction.

To diagonalize the Hamiltonian (\ref{eq:kp_Ham}), following Ref.~\cite{Kazakov:2025_PRB}, the basis states $|\lambda_j\rangle$ (Eq.~(\ref{eq:basis_phi})) are expanded in a truncated orthonormal plane-wave basis $\{| f_n \rangle\}$ of dimension $2N+1$:
\begin{equation}
|\lambda_j\rangle = \sum\limits_{n=-N}^{N}\alpha^{n}_j| f_n \rangle = \sum\limits_{n=-N}^{N} \frac{\alpha^{n}_j}{\sqrt{L_z}}\exp\left(\mathrm{i} \frac{2\pi n}{L_z}z \right) ,
\label{eq:fourierN}
\end{equation}
with $j\in\{ 1,2,3,4 \}$, $L_z = d_{\text{QW}}+2d_{\text{BR}}$ (Fig.~\ref{fig:QW}), and $\alpha^{n}_j$ the expansion coefficients. Consequently, each element $\hat{\mathcal{H}}_{ij}(\bm{\mathit{k}})$ becomes represented by a $(2N$~$+$~$1)$ $\times$ $(2N$~$+$~$1)$ matrix. One can then evaluate the matrix elements of the operator $\hat{k}_z^{\nu}$ $=$ $(-\mathrm{i})^{\nu}\partial_z^{\nu}$ $(\nu = 0, 1, 2)$, defined by $\mathcal{K}^{\nu}_{mn} = \langle f_m| \hat{k}_z^{\nu} |f_n\rangle$, namely
\begin{equation}
\mathcal{K}^{\nu}_{mn} = \left(\frac{2\pi n}{L_z}\right)^{\nu} \underbrace{\int\limits_{-L_z/2}^{L_z/2}\frac{\mathrm{d}z}{L_z}\ \exp\left(\mathrm{i}  \frac{2\pi}{L_z}(n-m)z \right)}_{\langle f_m|f_n\rangle} .
\label{eq:integral}
\end{equation}
Crucially, the integration domain must be partitioned into subregions governed by distinct parameter sets: the quantum well interior 
\begin{subequations}
\begin{equation}
z \in \left[-d_{\text{QW}}\, /\, 2,\, d_{\text{QW}}\, /\, 2\right],
\end{equation} 
and the barrier regions 
\begin{equation}
z \in \left[-L_z\, /\, 2,\, -d_{\text{QW}}\, /\, 2 \right[\ \cup\ \left] d_{\text{QW}}\, /\, 2,\, L_z\, /\, 2\right] .
\end{equation}
\end{subequations}
The contributions of these subdomains to the matrix $\mathcal{K}^{\nu}$ are denoted $\mathcal{K}^{\nu}_{\text{QW}}$ and $\mathcal{K}^{\nu}_{\text{BR}}$, respectively, such that 
\begin{equation}
\mathcal{K}^{\nu} = \mathcal{K}^{\nu}_{\text{QW}} + \mathcal{K}^{\nu}_{\text{BR}}\, .
\end{equation} 
After applying a suitable basis transformation (if necessary), the Hamiltonian $\hat{\mathcal{H}}(\bm{\mathit{k}})$ is assembled as a $4(2N$~$+$~$1)$ $\times$ $4(2N$~$+$~$1)$ matrix by replacing the differential operators $\hat{k}_z^{\nu}$ ($\hat{k}_z^{0}$ $\equiv$ $\mathrm{id}$) with the appropriate combinations of $\mathcal{K}^{\nu}_{\text{QW}}$ and $\mathcal{K}^{\nu}_{\text{BR}}$ determined by the parameter values multiplying each operator in the given subregion. Numerical diagonalization for a selected valley then yields eigenvalues and eigenvectors constituting the quantized energy levels and the carrier wavefunctions confined in the quantum well. 

\begin{figure*}[t]
	\centering
	
	\subfloat[\label{subfig:kafelki_a}]{\includegraphics[scale=1.0]{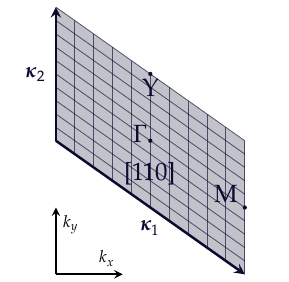}}\quad
    \subfloat[\label{subfig:kafelki_b}]{\includegraphics[scale=1.0]{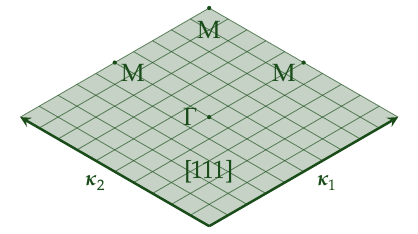}}\quad
    \subfloat[\label{subfig:kafelki_c}]{\includegraphics[scale=1.0]{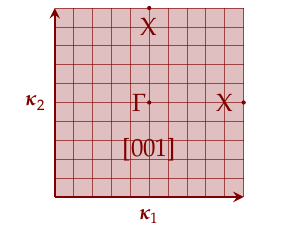}}
 \caption{Schematic diagrams of the discrete FBZ meshes for the  two-dimensional systems considered in this study (cf. Fig.~\ref{subfig:IV-VIs_b}). The spanning vectors $\bm{\mathit{\kappa}}_{\mu}$ are given by (a) $\bm{\mathit{\kappa}}_{1}=\frac{2\pi}{a_0}\left[\sqrt{2},-1\right]{}^{\mathrm{T}}$, $\bm{\mathit{\kappa}}_{2}=\frac{2\pi}{a_0}[0, 1]^{\mathrm{T}}$, (b) $\bm{\mathit{\kappa}}_{1}=\frac{2\pi}{a_0}\sqrt{\frac{2}{3}}\left[\sqrt{3}, 1\right]{}^{\mathrm{T}}$, $\bm{\mathit{\kappa}}_{2}=\frac{2\pi}{a_0}\sqrt{\frac{2}{3}}\left[-\sqrt{3}, 1\right]{}^{\mathrm{T}}$, (c) $\bm{\mathit{\kappa}}_{1}=\frac{2\sqrt{2}\pi}{a_0}[1, 0]^{\mathrm{T}}$, $\bm{\mathit{\kappa}}_{2}=\frac{2\sqrt{2}\pi}{a_0}[0, 1]^{\mathrm{T}}$, where $a_0$ is the fcc lattice constant.} 
\label{fig:kafelki}
\vspace{-0.0cm}
\end{figure*}
  
\subsection{Chern number calculations}
\label{subsec:Cherncalc}

The topological invariant $\mathcal{C}$ was computed numerically via the \textit{plaquette method} of Fukui \textit{et al.} \cite{Fukui:2005_JPhysSocJap}, wherein the Chern number is evaluated as the discrete sum of the Berry curvature $\mathcal{F}_{12}(\bm{\mathit{k}})$, $\bm{\mathit{k}} = [k_1,k_2]^{\mathrm{T}}$, over the quadrilateral plaquettes resulting from discretization of the two-dimensional FBZ (cf. Fig.~\ref{fig:kafelki}). A brief outline of this procedure is presented below to clarify its compatibility with the $\bm{\mathit{k}}\cdot\bm{\mathit{p}}$ framework, from which the electronic states $|\widetilde{\bm{\mathit{\alpha}}}_b(\bm{\mathit{k}})\rangle$ required for the implementation of the plaquette algorithm are obtained. Here $b$ is the subband index and 
\begin{equation}
\widetilde{\bm{\mathit{\alpha}}} \equiv \left[\bm{\mathit{\alpha}}_1^{\mathrm{T}}, \bm{\mathit{\alpha}}_2^{\mathrm{T}}, \bm{\mathit{\alpha}}_3^{\mathrm{T}}, \bm{\mathit{\alpha}}_4^{\mathrm{T}}\right]^{\mathrm{T}},
\end{equation}
with $\bm{\mathit{\alpha}}_j$ denoting the vector of coefficients $\mathit{\alpha}^{n}_j$ appearing in Eq.~(\ref{eq:fourierN}) (subband index and $\bm{\mathit{k}}$-dependence are omitted for brevity). In order to include contributions from all $\widetilde{N}=2(2N+1)$ occupied valence subbands (with possible degeneracies), the eigenvectors of the Hamiltonian are arranged as columns of the matrix
\begin{equation}
\hat{\Upsilon}(\bm{\mathit{k}}) \equiv \left[ |\widetilde{\bm{\mathit{\alpha}}}_{1}(\bm{\mathit{k}})\rangle, |\widetilde{\bm{\mathit{\alpha}}}_{2}(\bm{\mathit{k}})\rangle, \dots, |\widetilde{\bm{\mathit{\alpha}}}_{\widetilde{N}}(\bm{\mathit{k}})\rangle \right] .
\label{eq:Upsilon}
\end{equation}
The unitary overlap (link) matrices are then defined by
\begin{equation}
\hat{U}_{\mu}(\bm{\mathit{k}}) \equiv \hat{\Upsilon}^{\dagger}(\bm{\mathit{k}})\hat{\Upsilon}(\bm{\mathit{k}} + \Delta\bm{\mathit{k}}_{\mu})\, ,
\label{eq:link}
\end{equation}
where $\Delta\bm{\mathit{k}}_{\mu}$ is the grid spacing in direction $\mu\in\{ 1,2\}$. The discrete analogue of the \textit{Berry connection} $\mathcal{A}_{\mu}(\bm{\mathit{k}})$, measuring the net phase difference $(\mathrm{mod}\ 2\pi)$ of the eigenvectors at neighbouring nodes $\bm{\mathit{k}}$ and $\bm{\mathit{k}} + \Delta\bm{\mathit{k}}_{\mu}$, is given~by
\begin{equation}
\mathcal{A}_{\mu}(\bm{\mathit{k}}) \equiv \mathrm{i}\, \mathrm{Im} \left[\ln \det\left( \hat{U}_{\mu}(\bm{\mathit{k}})\right)\right] ,
\label{eq:BConV}
\end{equation}
and the \textit{Berry curvature} takes the form
\begin{subequations}
\begin{align}
\mathcal{F}_{12}(\bm{\mathit{k}}) & \equiv \mathrm{i}\, \mathrm{Im} \left[\ln \det\left( \hat{U}_{1}(\bm{\mathit{k}})\hat{U}_{2}(\bm{\mathit{k}} + \Delta\bm{\mathit{k}}_1)\right.\right. \nonumber \\ 
&\ \ \ \ \ \ \ \ \ \ \ \ \ \ \ \ \ \ \ \ \ \times \left.\left.\hat{U}_{1}^{\dagger}(\bm{\mathit{k}} + \Delta\bm{\mathit{k}}_2)\hat{U}_{2}^{\dagger}(\bm{\mathit{k}})\right)\right]
\label{eq:BDFukui_a}
\end{align}
or equivalently
\begin{equation}
\mathcal{F}_{12}(\bm{\mathit{k}}) = \Delta_{1}\mathcal{A}_{2}(\bm{\mathit{k}}) - \Delta_{2}\mathcal{A}_{1}(\bm{\mathit{k}}) + 2\pi\mathrm{i}\, n(\bm{\mathit{k}})\, ,
\label{eq:BDFukui_b}
\end{equation}
\end{subequations}
with $\Delta_{\mu}\mathcal{A}_{\nu}(\bm{\mathit{k}}) \equiv \mathcal{A}_{\nu}(\bm{\mathit{k}} + \Delta\bm{\mathit{k}}_{\mu}) - \mathcal{A}_{\nu}(\bm{\mathit{k}})$, and $n(\bm{\mathit{k}})$~$\in$~$\mathbb{Z}$ an integer-valued field such that $-\pi<\mathrm{Im}\left[\mathcal{F}_{12}(\bm{\mathit{k}})\right] \leqslant \pi$. The contributions coming from the differences $\Delta_{\mu}\mathcal{A}_{\nu}(\bm{\mathit{k}})$ associated with adjacent plaquettes cancel pairwise; consequently the Chern number
\begin{equation}
\mathcal{C} = \frac{1}{2\pi\mathrm{i}}\sum\limits_{\bm{\mathit{k}}}\mathcal{F}_{12}(\bm{\mathit{k}}) = \sum\limits_{\bm{\mathit{k}}}n(\bm{\mathit{k}})
\end{equation}
is manifestly an integer. Importantly, the plaquette algorithm is invariant under the local phase transformation
\begin{equation}
|\widetilde{\bm{\mathit{\alpha}}}_b(\bm{\mathit{k}})\rangle \mapsto \exp\left(\mathrm{i}\varphi_b(\bm{\mathit{k}})\right)|\widetilde{\bm{\mathit{\alpha}}}_b(\bm{\mathit{k}})\rangle
\end{equation}
and remains valid on coarser numerical meshes. 

Although the $\bm{\mathit{k}}\cdot\bm{\mathit{p}}$ approximation is formally justified only in a small neighbourhood of $\bm{\mathit{k}} = 0$ (which coincides with the valley point), it is nevertheless adequate for employing the plaquette-based evaluation of the Chern invariant in the quantum wells studied here. For these systems the value of the topological invariant is determined solely by the (non)trivial band ordering at the gap point, and the $\bm{\mathit{k}}\cdot\bm{\mathit{p}}$ description reliably reproduces the critical portion of the FBZ which governs the topology. Moreover, the nonperiodicity of the Hamiltonian in reciprocal space is immaterial here, because contributions to the Chern number from regions distant from the band extrema are negligible. However, the topological invariant must be computed separately for each valley type, and the Chern number for a given quantum well deduced from the resulting valley-resolved values.

\begin{figure}[b!]
	\centering
	\includegraphics[width=0.91\columnwidth]{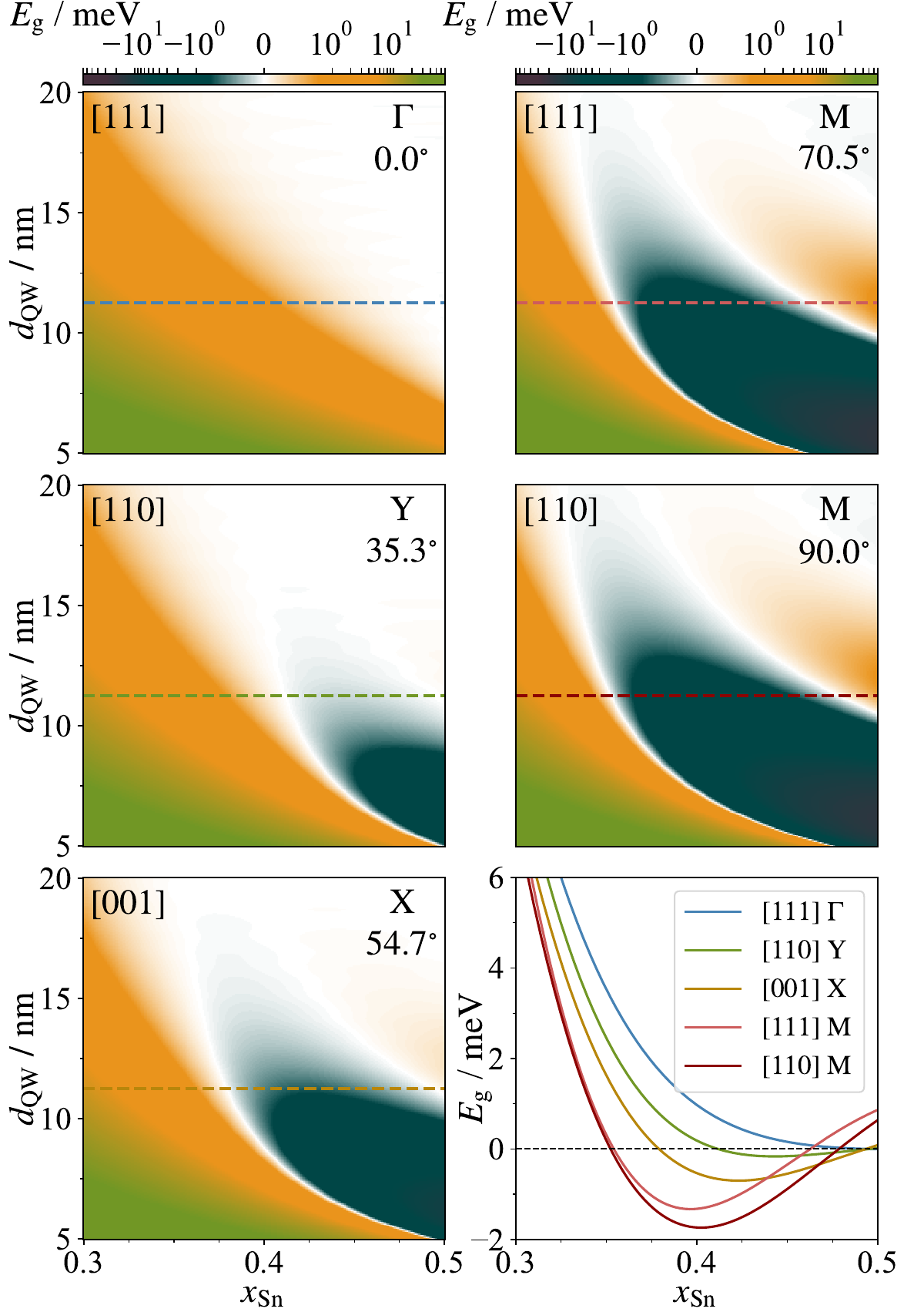}	
	\vspace{-0.0cm}
 \caption{Band gap diagrams as functions of tin concentration $x_{\text{Sn}}$ and the quantum well thickness $d_{\text{QW}}$ for the individual valleys of structures grown along $[hkl]$. The value $\vartheta$ beneath the symbol for a given $\mathrm{L}$-valley projection indicates the three-dimensional angular deviation of the valley axis from the well growth direction (cf. Fig.~\ref{fig:TR}). Other system parameters are $B = 0.0\, \mathrm{T}$, $T = 1.5\,\mathrm{K}$, $d_{\text{BR}} = 15\,\mathrm{nm}$, $y_{\text{EuS}} = 0.25$. The lower-right panel displays cross sections of the maps $E_{\text{g}}(x_{\text{Sn}}, d_{\text{QW}})$ taken along the dashed lines at $d_{\text{QW}} = 11.25\, \mathrm{nm}$.} 
\label{fig:EgDiag}
\end{figure}

\begin{figure*}[t!]	
	\centering
		\includegraphics[width=\textwidth]{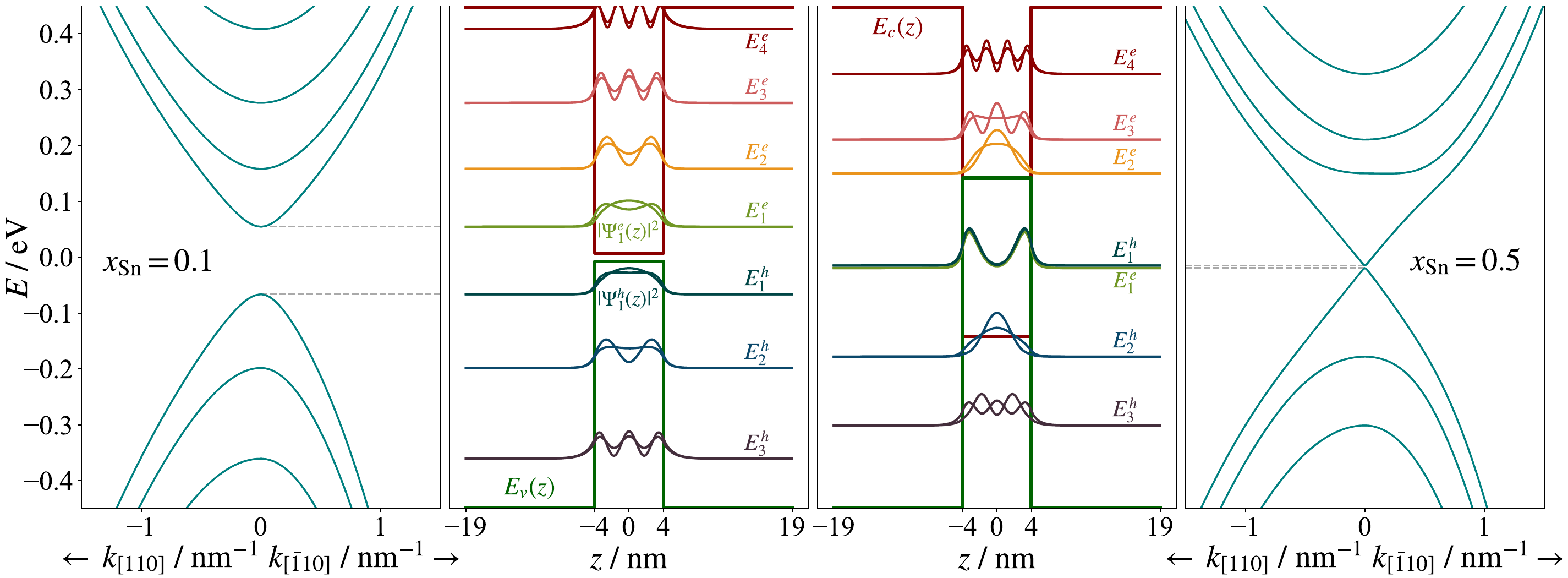}
\vspace{-0.5cm}	
	\caption{The calculated band dispersions $E_{b}^{e,h}(\bm{\mathit{k}})$ and the associated probability densities $|\Psi_{b}^{e,h}(z)|^2$ (arbitrary units) for band-edge states of a $[001]$-oriented quantum well, presented for the trivial $(x_{\text{Sn}}=0.1)$ and nontrivial $(x_{\text{Sn}}=0.5)$ topological phases. Structure parameters are as in Fig.~\ref{fig:EgDiag}. The central panels display the confining potential profiles that define the band edges, \textcolor{darkred}{$E_c(z)$} and \textcolor{trawa}{$E_v(z)$}, in the respective segments of the heterostructure. Attention is drawn to the spatial localization of surface states adjacent to the well boundaries in the inverted-band regime and on the gapped Dirac-cone (nearly linear) dispersion exhibited by these states.}
	\label{fig:QWF}
\end{figure*}

\section{Results}
\label{sec:results}

\subsection{Quantum well band structures}
\label{subsec:QWBands}

Numerical calculations underlying all the results presented below rely on $N = 60$ Fourier components in Eq.~(\ref{eq:fourierN}), corresponding to a $484\times 484$ Hamiltonian matrix; this truncation provides satisfactory convergence of the computed quantities.

In order to identify regions in the structural-parameter space favorable for the QAH phase, the band gap, defined as $E_{\text{g}}\equiv (E_1^{e}-E_1^h)(\bm{\mathit{k}}=0)$ (cf. Fig.~\ref{fig:QW}), is computed assuming complete demagnetization of the magnetic ions in the barrier layers. In that case, the QAH regime becomes accessible whenever $E_{\text{g}} < 0$. Subsequent magnetization of the magnetic dopants produces an exchange splitting that lifts the spin degeneracy so that the band inversion may be eliminated for one spin channel but retained for the other, thereby yielding a single inverted spin-resolved band pair necessary for the QAHE. Representative color maps $E_{\text{g}}(x_{\text{Sn}}, d_{\text{QW}})$ for all examined growth directions $[hkl] \parallel \hat{\bm{\mathit{e}}}_z$ and corresponding valley projections are provided in Fig.~\ref{fig:EgDiag}. Two topologically distinct phases --- distinguished by the sign of the band gap --- appear for each quantum well orientation. There exists a clear correlation between the angle $\vartheta$ (see the caption of Fig.~\ref{fig:EgDiag}) and the composition at which the first inversion of the $E^{e}_{1}$ and $E^{h}_{1}$ levels occurs upon varying $x_{\text{Sn}}$ at fixed $d_{\text{QW}}$. The anisotropy of the constant-energy surfaces (cf. $m_t < m_l$ and $v_t > v_l$ in Table~\ref{tab:kp_params}) favors larger angular deviations of the valley axis; accordingly, at $\vartheta = 90${\textdegree} the gap sign reversal takes place at the lowest tin concentration, while progressively higher concentrations are required for valleys at smaller deviation angles. In particular, no band gap inversion is observed for the $\Gamma$ valley $(\vartheta = 0${\textdegree} relative to $[111])$ within the model’s range of applicability.
 
Another notable feature is a damped oscillatory dependence of the band gap on both $d_{\text{QW}}$ and $x_{\text{Sn}}$, arising from hybridization of states localized at the two interfaces separating materials of distinct topological character. In the former case the overlap of these states is governed by the distance between them, whereas in the latter it is controlled by the depth of the confining potential. A comparable finding for the $\mathrm{M}$ valley in $[111]$-oriented wells was reported in Ref.~\cite{Kazakov:2025_PRB}, albeit using a different effective description of the states at that point. Furthermore, the maps in Fig.~\ref{fig:EgDiag} suggest that $[111]$ and $[110]$ quantum wells, with the barrier layers demagnetized (thus preserving $\mathrm{TRS}$), can realise the QSHE, since these systems exhibit regions where the gap is inverted at an odd number of points in the two-dimensional $\mathrm{FBZ}$. 

\begin{figure*}[t]	
	\begin{center}
		\includegraphics[width=0.95\textwidth]{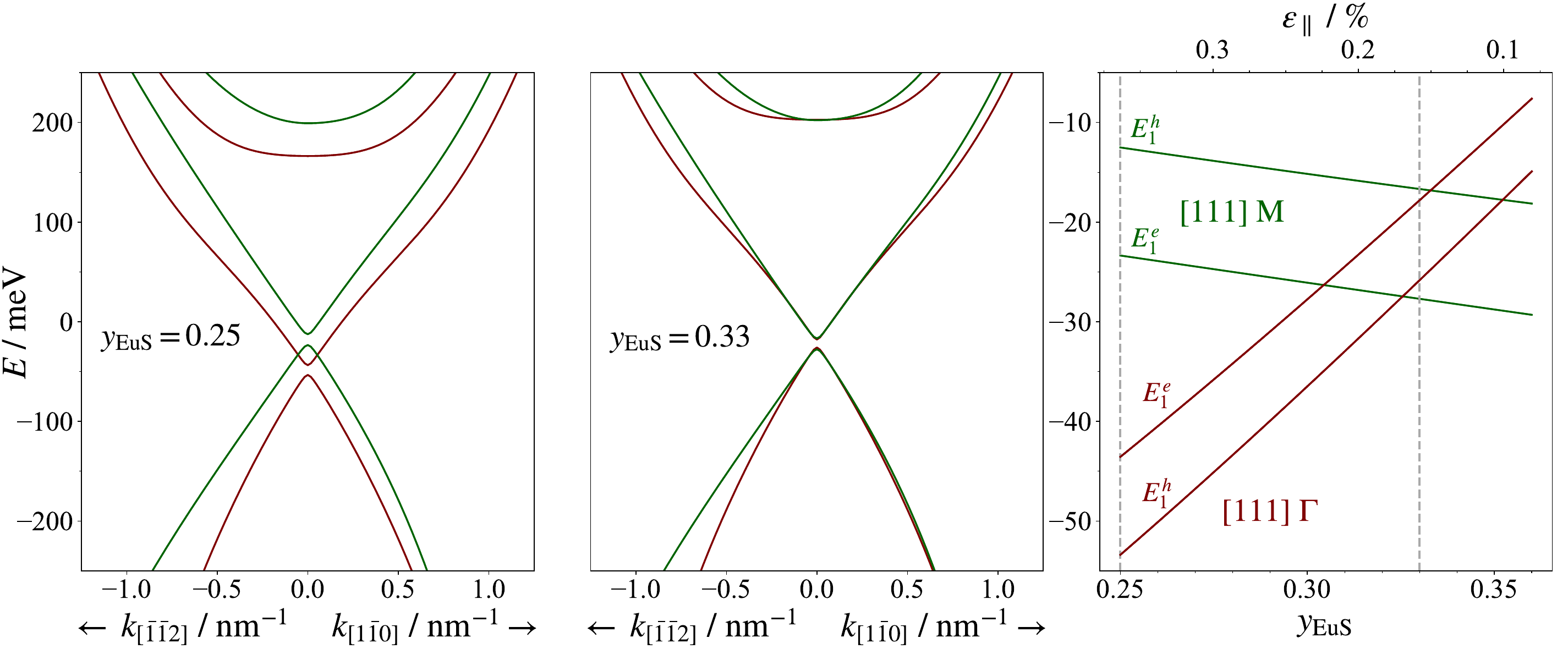}
	\end{center}\vspace{-0.5cm}	
	\caption{Strain-induced control of the relative band-edge alignment in the $\Gamma$ and $\mathrm{M}$ valleys of a $[111]$-oriented quantum well. Results shown correspond to $x_{\text{Sn}} = 0.49$ and $d_{\text{QW}} = 5\, \mathrm{nm}$; all other parameters follow Fig.~\ref{fig:EgDiag}. At $y_{\text{EuS}} = 0.25$, states originating from the $\Gamma$ valley reside within the topological gaps of the three M valleys. A reduction of the lattice mismatch between the well and the barriers --- measured by the in-plane strain component $\varepsilon_{\parallel}$ --- produces systematic shifts of the band edges in both valley families. By appropriately tuning $\varepsilon_{\parallel}$ (without changing the intrinsic level ordering) the longitudinal-valley states can be expelled from the spectral window that constitutes the gap in the oblique valleys. For clarity, the magnetic dopants are assumed to be demagnetized here; however, the mechanism presented remains effective when the spin degeneracy is lifted.}
	\label{fig:strain}
	\vspace{-0.0cm}
\end{figure*} 

 \begin{figure}[b!]
 \centering	
		\includegraphics[width=1.0\columnwidth]{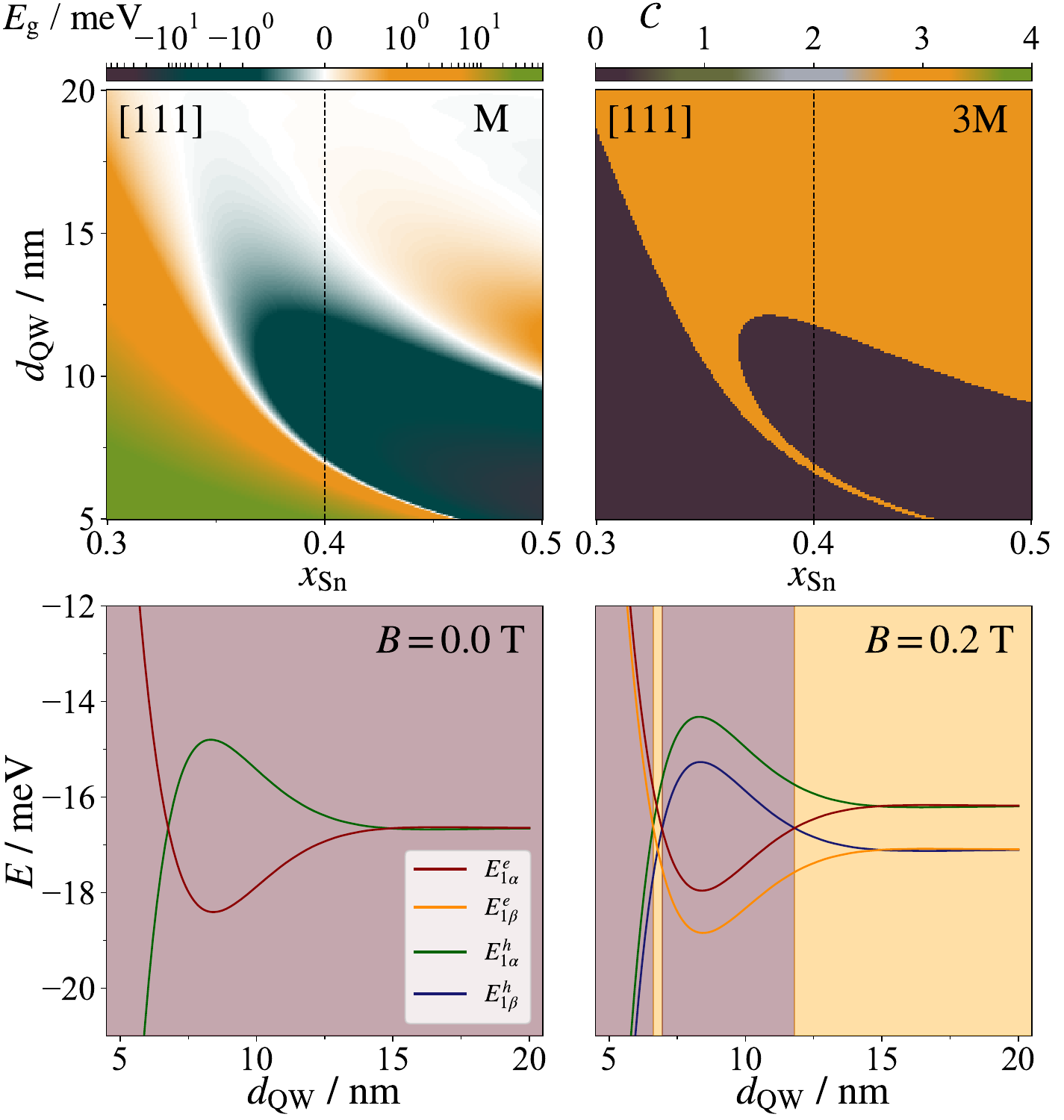}
		\vspace{-0.25cm}
	\caption{Spin splitting of the bands at $\bm{\mathit{k}} = 0$ in a magnetic field $\bm{\mathit{B}} = B\hat{\bm{\mathit{e}}}_z$, illustrated for the $\mathrm{M}$ valley of a $[111]$ quantum well. Top panels show (from left) a map of the band gap ($B=0\, \mathrm{T}$) and the Chern number phase diagram ($B=0.2\, \mathrm{T}$). Bottom panels compare the corresponding spin-resolved levels $E^{h,e}_{1\alpha}$ and $E^{h,e}_{1\beta}$ along the dashed lines indicated in the top panels.}
	\label{fig:splitting}
\end{figure} 

Figure~\ref{fig:QWF} presents band dispersions and the corresponding carrier wavefunctions for a $[001]$-oriented quantum well, shown for cases with differing signs of the band gap $E_{\text{g}}$. For the trivial level ordering ($x_{\text{Sn}} = 0.1$) a conventional parabolic dispersion is obtained, slightly deformed by anisotropy of the $\bm{\mathit{k}} \cdot \bm{\mathit{p}}$ parameters. In the inverted regime --- characterized by the interchange of the $E_1^h$ and $E_1^e$ levels at $\bm{\mathit{k}} = 0$ ($\mathrm{X}$) --- the associated bands form a massive Dirac cone ($E_{\text{g}} \neq 0$); a representative cross section for $x_{\text{Sn}} = 0.5$ is displayed in one panel of the Fig.~\ref{fig:QWF}. The probability density profiles $|\Psi_{b}^{e,h}(z)|^2$, plotted at the nominal band-edge energies, constitute a direct manifestation of the system’s topological character (specifically, a 2D TCI phase). The wavefunctions $\Psi_{b}^{e,h}(z)$ are constructed from Eq.~(\ref{eq:fourierN}) by summing the relevant Fourier components with coefficients $\alpha_j^{n}$ produced by Hamiltonian diagonalization. A clear signature of gapless surface states tied to the $E_1^h$ and $E_1^e$ levels is the pronounced localization of $|\Psi_{1}^{h}(z)|^2$ and $|\Psi_{1}^{e}(z)|^2$ at the quantum well edges, which implies that carrier transport occurs predominantly along the interfaces between the topological inner layer and the trivial barriers.

\begin{figure*}[t!]	
	\centering
		\includegraphics[width=0.9\textwidth]{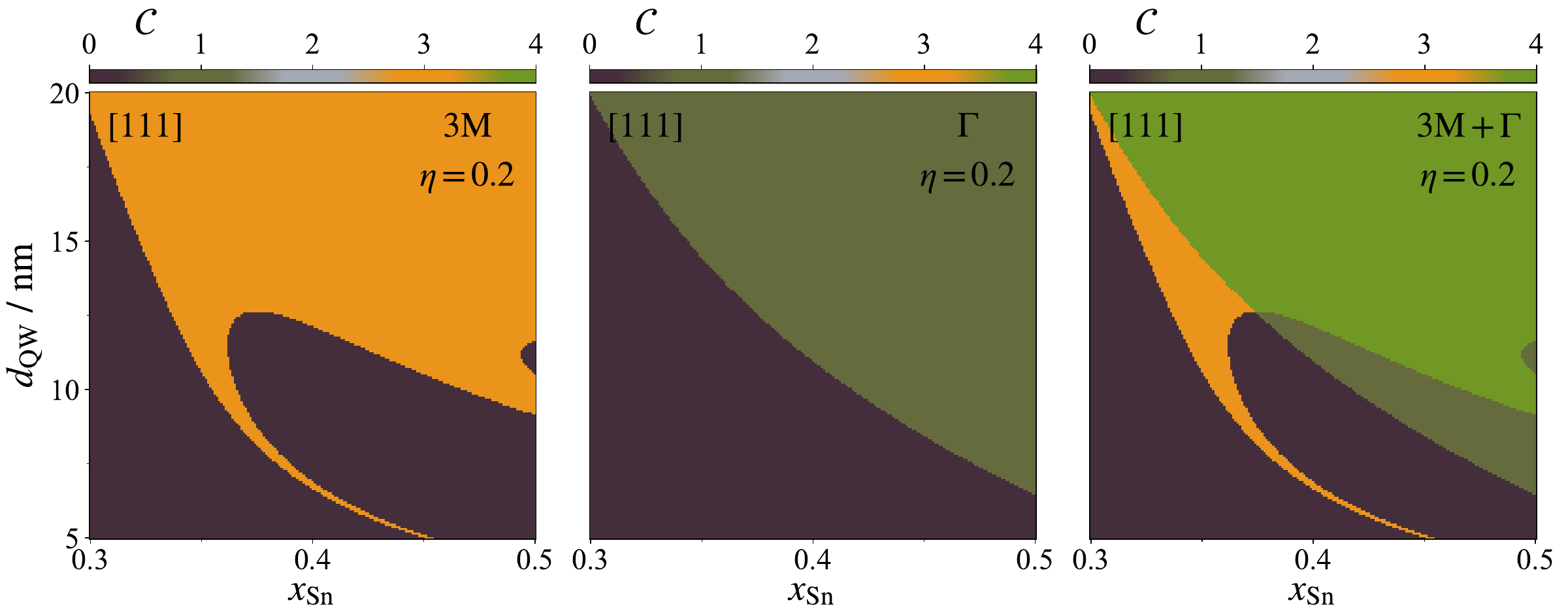}
	\vspace{-0.2cm}	
	\caption{Procedure for assembling the Chern number phase diagram, shown for a $[111]$ quantum well with assumed barrier magnetization coefficient $\eta = 0.2$ (other parameters as in Fig.~\ref{fig:EgDiag}). The rightmost panel results from superimposing the Chern number values evaluated at each point for all projections of the $\mathrm{L}$ valleys onto the two-dimensional $\mathrm{FBZ}$. All computations of the topological invariant were performed on a $13\times 13$ numerical grid.}
	\label{fig:sumlo}
	\vspace{-0.2cm}
\end{figure*}

From the standpoint of realizable nanostructures, the ability to account for strain is especially valuable, since it supplies an additional degree of freedom for controlling the system’s properties. In the present context, ensuring accurate quantization of the transverse conductance in the QAHE critically requires that the topological gap contain no other states (aside from the edge states at the well boundaries) that could give rise to bulk conduction. The multivalley character of IV--VI materials may, however, present challenges in this respect: the splitting of the four projected $\mathrm{L}$ valleys into two inequivalent subsets, combined with strain effects and the anisotropy of isoenergetic surfaces near the band edges, can result in states associated with one valley populating the spectral gap of another --- a situation observed for $[111]$- and $[110]$-oriented wells. Crucially, variation of the EuS fraction in the barrier layers provides a practical route to tune the lattice mismatch between constituent layers and thus to shift the relative band-edge positions across valleys, enabling the elimination of unwanted in-gap states. The mechanism is exemplified in Fig.~\ref{fig:strain} for the $\Gamma$ and $\mathrm{M}$ valleys of a $[111]$ well. It should be emphasised that employing EuSe (which is more naturally lattice-matched to PbSe) as the barrier material instead of EuS would permit only tensile in-plane strain ($\varepsilon_{\parallel} > 0$; cf. Eq.~(\ref{eq:epsy})), since alloying PbSe with Sn reduces the well lattice constant and $a_{\text{EuSe}} > a_{\text{PbSe}}$. In contrast, EuS has $a_{\text{EuS}} < a_{\text{PbSe}}$, which offers increased tunability of the heterostructure strain and, in particular, allows also realization of compressive in-plane strain $(\varepsilon_{\parallel} < 0)$. Finally, the above complications play a substantially reduced role for $[001]$ quantum wells, in which all projections of the $\mathrm{L}$ valleys are equivalent. Yet tensile strain still shifts the band edges --- i.e., it increases the trivial gap --- and therefore should be minimized. 
 
\subsection{Chern number phase diagrams}
\label{subsec:ChNPhD}

Having identified regions of the structural-parameter space that are favorable for the QAHE, the subsequent task is to investigate the effect of TRS breaking on the topology of the calculated band structures. Depending on the magnetic-dopant content $y_{\text{EuS}}$ in the barriers, TRS may be broken either by application of a small magnetic field that facilitates ordering of the magnetic moments, or --- at sufficiently high dopant concentrations --- by introducing a finite barrier magnetization parameter $\eta$ (cf. the discussion following Eq.~(\ref{eq:Delta_dsf})). For the value $y_{\text{EuS}} = 0.25$ used in the present calculations both strategies are equally justified; illustrative results for each approach are given below.

The results in Fig.~\ref{fig:splitting} demonstrate the mechanism by which magnetization of the barrier layers drives the system into the QAHE regime, shown here for the $\mathrm{M}$ valley of a $[111]$-oriented quantum well (parameters as in Fig.~\ref{fig:EgDiag}). In the absence of a magnetic field the band levels remain spin-degenerate across the parameter space, yielding $\mathcal{C} = 0$. For magnetized Eu spins in the barriers $(B\neq 0)$ $\mathcal{C} \neq 0$ emerge near $E_{\text{g}} \approx 0$ once the induced spin splitting attains a sufficient magnitude. Importantly, the intervals of well thickness $d_{\text{QW}}$ that exhibit nonzero Chern number coincide with those for which the level inversion involves only one spin-resolved pair while the complementary pair remains uninverted. As a result, and because the band gap oscillates at $B = 0$, islands of the ordinary-insulator phase persist on the $\mathcal{C}(x_{\text{Sn}}, d_{\text{QW}})$ phase maps in regions where $|E_{\text{g}}(B=0)|$ still exceeds the exchange-induced spin splitting.

\begin{figure*}[t!]	
	\centering
		\includegraphics[width=0.9\textwidth]{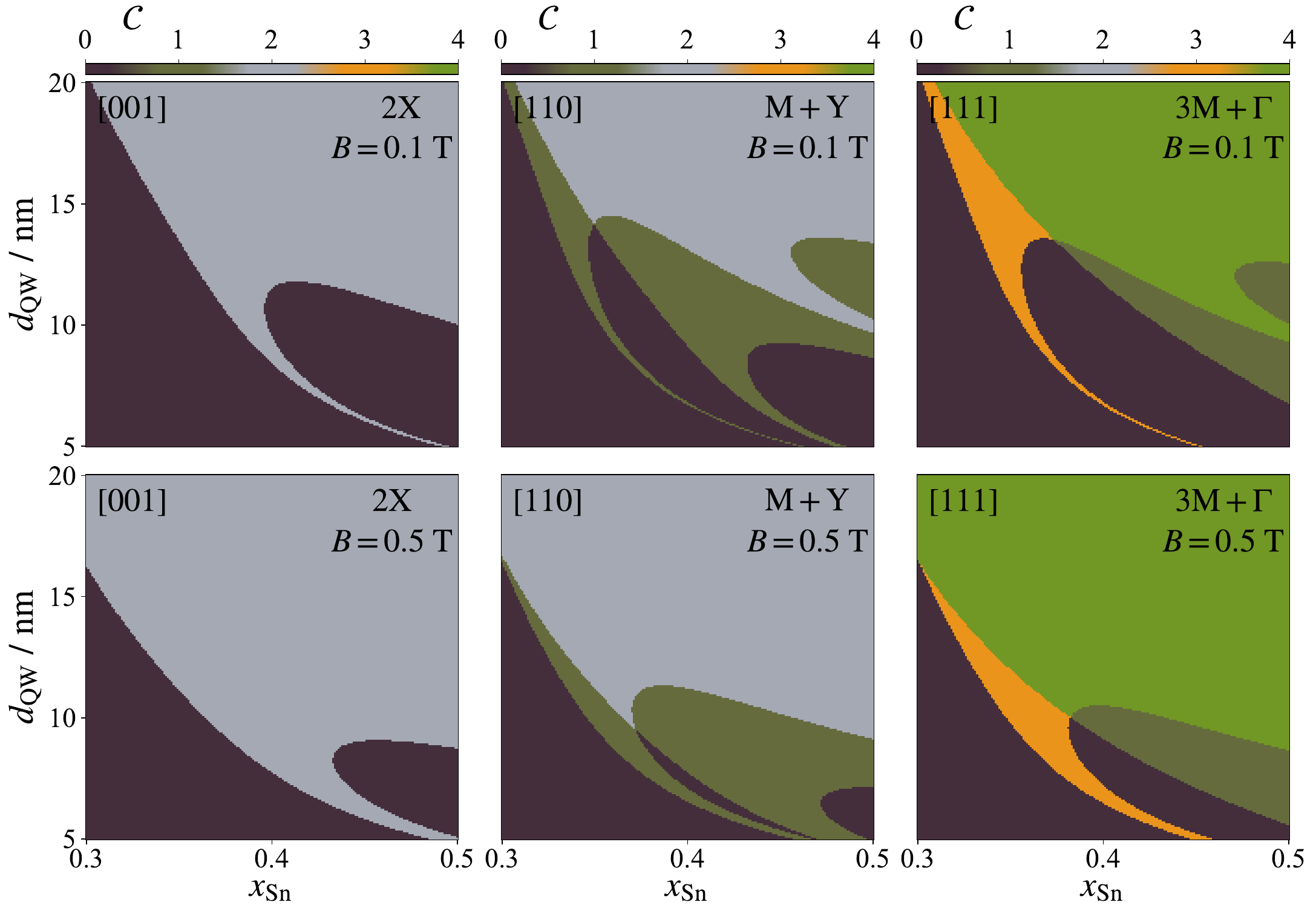}
	\vspace{-0.2cm}	
	\caption{Compilation of Chern number phase diagrams for the three considered quantum well growth directions $[hkl]$. For each orientation, series of diagrams computed for two values of the external magnetic field $\bm{\mathit{B}} = B\hat{\bm{\mathit{e}}}_z$ are presented (remaining parameters as in Fig.~\ref{fig:EgDiag}), following the procedure of Fig.~\ref{fig:sumlo} that accounts for all valley projections within the same heterostructure. Appropriate choice of the magnetization coefficient $\eta$ reproduces qualitatively equivalent phase patterns, so the $B=0$ case can be omitted.}
	\label{fig:ChernPhD}
	\vspace{-0.2cm}
\end{figure*}

In the phase diagram of Fig.~\ref{fig:splitting}, the Chern number takes the value $\mathcal{C} = 3$ within the QAH phase. This value reflects that, for the $[111]$ orientation, three \textit{distinct} $\mathrm{M}$ valleys exist that nevertheless exhibit \textit{identical} energetic structure --- each valley contributing at most $|\mathcal{C}| = 1$ to the total invariant. A complete phase description for the well requires inclusion of the $\Gamma$ valley as well, which likewise is capable of producing $\mathcal{C} \neq 0$, consistent with the very small band gap $E_{\text{g}}$ displayed in Fig.~\ref{fig:EgDiag} at elevated tin fractions and larger well thicknesses. Figure~\ref{fig:sumlo} provides an illustration by juxtaposing representative Chern number maps for the different valley types of the $[111]$ structure and their superposition. The calculations assume a barrier magnetization coefficient $\eta = 0.2$; consequently, $\mathcal{C} \neq 0$ in this context indicates the quantum anomalous Hall effect in the strict sense (i.e., occurring without an external magnetic field). These results further highlight the wealth of QAH phases afforded by IV--VI semiconductors, a property that derives from the placement of the band gap at the $\mathrm{L}$ points. In the present case the largest possible Chern number is $\mathcal{C} = 4$ (note that $|\mathcal{C}| = 2$ does not occur), whereas for the other growth directions the maximal value is $\mathcal{C} = 2$ (see below).

Figure~\ref{fig:ChernPhD} depicts representative Chern number phase diagrams for each quantum well variant considered, each shown for two values of the applied magnetic field. As can be seen, nontrivial regions --- characterised by a nonzero topological invariant --- expand as larger fields induce stronger barrier magnetization, encroaching on areas of trivial topology. These results demonstrate that, within the parameter ranges investigated, the modeled (Pb,Sn)Se-based heterostructures can support multiple QAH phases. Accordingly, three additional classes of heterostructures are identified as promising candidates for experimental exploration of topological phases, including QAH states with Chern number $\mathcal{C} > 1$. It should be emphasized that the present analysis is restricted to barrier magnetization along $\hat{\bm{\mathit{e}}}_z$, since this orientation enhances the width of the topological gap once spin degeneracy is lifted and enables direct comparison of the two TRS-breaking scenarios considered. An in-plane magnetization component mixes the Kramers-pair states $\alpha$ and $\beta$ (for $\bm{\mathit{b}}\parallel \hat{\bm{\mathit{e}}}_z$ this coupling, if present, remains small), and, when $\mathcal{C} \neq 0$, typically causes the topological gap to shrink at some finite $\bm{\mathit{k}}$ away from the valley point.

The simplifications inherent in the $\bm{\mathit{k}}\cdot\bm{\mathit{p}}$ description imply that the phase boundaries shown in Fig.~\ref{fig:ChernPhD} should not be regarded as strictly exact but as indicative. Nonetheless, they provide a robust foundation for device design, and experimental investigation of structures based on these predictions would provide the ultimate validation of both the modelling framework and the conclusions drawn here.

\section{Summary and outlook}
\label{subsec:SumOut}

To conclude, this study supplies practical methodologies for probing the topological character of (Pb,Sn)Se quantum wells with magnetic barriers and marks an important advance toward the fabrication of real nanostructures that exhibit the QAHE on an IV--VI materials platform. Theoretical analysis demonstrates that wells grown along the $[001]$, $[110]$ and $[111]$ orientations are capable of supporting quantized Hall conductance over wide ranges of structural and compositional parameters. Owing to the engineering focus of the modelling approach, the obtained results can be readily transferred to experimental implementation. Of particular note is the potential realization of QAH phases with $\mathcal{C} > 1$, to date, realized only in multiple quantum well systems \cite{Zhao:2020_Nature}. 

To the best of our knowledge, this work constitutes the first instance of modeling the electronic structure of systems based on this class of semiconductors within the $\bm{\mathit{k}} \cdot \bm{\mathit{p}}$ framework for a choice of the growth axis $\hat{\bm{\mathit{e}}}_z$ other than $[111]$. The proposed basis transformation of the Hamiltonian (\ref{eq:kp_Ham}) affords, with relatively low computational expense, access to the energy spectra corresponding to projections of the $\mathrm{L}$ valleys other than the $\Gamma$ $([111])$ projection, while rigorously including the anisotropy of the material’s constant-energy surfaces. A principal restriction of the approach is the limited availability of empirical parameters for selenides as compared to tellurides. Nevertheless, the framework captures the principal features of topological character of the investigated quantum wells and provide approximate guidance as to the parameter regimes in which particular QAH phases are attainable.

In view of the results reported here, $[001]$-oriented QWs offer the most favourable route to the QAHE, due to the equivalence of the $\mathrm{X}$ valleys. However, careful compensation of strain in the heterostructure remains essential to render the topological phases accessible at the lowest practicable Sn concentrations. By contrast, for the $[110]$ and $[111]$ orientations the experimental realization may require precise adjustment of the relative band-edge positions across the various valleys. In such cases the integration of a piezoelectric layer into the heterostructure --- permitting in situ tuning of strain via an applied voltage --- would allow the alignment of otherwise mismatched band edges and thus facilitate the realization of QAH phases with elevated Chern numbers, in particular $\mathcal{C} = 3$ and $\mathcal{C} = 4$ in $[111]$ systems.

We hope that the theoretical predictions put forward in this work will soon be subjected to experimental scrutiny, allowing for a realistic appraisal of the practical prospects of the considered architectures and clarifying the position of IV--VI materials within the broader landscape of QAHE platforms --- and, when coupled to superconductors, potentially also of topological superconductors.

\begin{acknowledgments}
We thank Alexander Kazakov and Valentine Volobuev for valuable discussions. This research was supported by the Foundation for Polish Science project "MagTop" no. FENG.02.01-IP.05-0028/23 co-financed by the European Union from the funds of Priority 2 of the European Funds for a Smart Economy Program 2021-2027 (FENG). Publication subsidized from the state budget within the framework of the programme of the Minister of Science (Polska) called Polish Metrology II project no. PM-II/SP/0012/2024/02.
\end{acknowledgments}

\section*{DATA AVAILABILITY}
The data that support the findings of this article are openly available \cite{Zenodo}.

\appendix

\section{Strain modelling}
\label{sec:appendix_A}

This work treats strain within the $\bm{\mathit{k}}\cdot\bm{\mathit{p}}$ formalism using deformation-potential theory. The strain-induced shift of the energy of the $m$-th valley in band $b$, evaluated to first order in perturbation theory, reads \cite{Singleton:1986_JPC_SSP} 
\begin{equation}
\delta E_{m}^{b} = (D_{m}^{b})^{ij}\varepsilon_{ij}\, ,
\label{eq:shift}
\end{equation} 
where $D_{m}^{b}$ denotes the deformation-potential tensor and $\varepsilon$ the strain tensor. The Einstein summation convention over repeated indices is adopted with the correspondence $\{x\leftrightarrow 1, y \leftrightarrow 2, z \leftrightarrow 3\}$. The analysis begins with the $\hat{\bm{\mathit{e}}}_z \parallel [111]$ case and its results are then invoked in the subsequent discussion of the remaining growth directions $[110]$ and $[001]$.

Lattice mismatch between the well material ($a_{\text{QW}}$) and the barriers ($a_{\text{BR}}$) induces an in-plane \textit{dilatational} strain in the $(111)$ interface, accompanied by a \textit{uniaxial} strain along the $[111]$ growth direction. The associated energy shifts are described by the deformation-potential constants $\Xi_{d}^{b}$ and $\Xi_{u}^{b}$, corresponding to the dilatational and uniaxial components, respectively, where $b$ labels the electronic band. 

The strain tensor is diagonal in the coordinate system aligned with the geometry of the problem, for which the axes were taken as $\left\{ [\bar{1}\bar{1}2], [1\bar{1}0], [111]\right\}$ (cf. Fig.~\ref{fig:TR}). In this basis one obtains
\begin{subequations}
\label{eq:epsy}
\begin{equation}
\varepsilon^{\prime} = \mathrm{diag}\left( \varepsilon_{\parallel},\varepsilon_{\parallel},\varepsilon_{\perp} \right),
\end{equation}
with
\begin{equation}
\varepsilon_{\parallel} = \frac{a_{\text{BR}} - a_{\text{QW}}}{a_{\text{QW}}} = \frac{a_{\text{BR}}}{a_{\text{QW}}} - 1\, , \\
\label{eq:eps_par}
\end{equation}
\begin{equation}
\varepsilon_{\perp} = -2\frac{C_{11} + 2C_{12} - 2C_{44}}{C_{11} + 2C_{12} + 4C_{44}}\varepsilon_{\parallel}\, .
\label{eq:eps_per}
\end{equation}
\end{subequations}
Here $\varepsilon_{\parallel}$ denotes the in-plane strain of the quantum well (the barrier lattice constants are taken as fixed), while $\varepsilon_{\perp}$ is the strain component normal to the plane. The relation between $\varepsilon_{\parallel}$ and $\varepsilon_{\perp}$ is derived from the stiffness (elasticity) tensor $C$ under the condition of vanishing stress in the $[111]$ direction. In the framework of linear elasticity the stress ($\sigma$) and strain ($\varepsilon$) tensors are related by 
\begin{equation}
\sigma_{ij} = C_{ij}{}^{kl}\varepsilon_{kl}\, .
\end{equation}
In the geometry-aligned basis the stress tensor is also diagonal, i.e., $\sigma^{\prime}=\mathrm{diag}\left( \sigma_{11}, \sigma_{22}, \sigma_{33} \right)$, with $\sigma_{11}=\sigma_{22}$ and $\sigma_{33}=0$. Hence
\[
0=\sigma_{33}=C^{\prime}_{33}{}^{ij}\varepsilon^{\prime}_{ij} = \left( C^{\prime}_{33}{}^{11} + C^{\prime}_{33}{}^{22} \right)\varepsilon_{\parallel} + C^{\prime}_{33}{}^{33}\varepsilon_{\perp}
\]
\begin{equation}
\Rightarrow \varepsilon_{\perp} = -\frac{C^{\prime}_{33}{}^{11} + C^{\prime}_{33}{}^{22}}{C^{\prime}_{33}{}^{33}}\varepsilon_{\parallel}\, .
\label{eq:stress}
\end{equation}
The number of independent components of the stiffness tensor depends on the symmetry of the crystal lattice. For cubic systems there are three independent elastic constants, $C_{11}$, $C_{12}$ and $C_{44}$. In the standard basis $\left\{ [100], [010], [001] \right\}$ the tensor $C$ takes the form
\begin{subequations}
\begin{equation}
C_{ij}{}^{kl} = \left\{ \begin{array}{cc}
C_{11} & i=j=k=l , \\
C_{12} & i=j\neq k=l , \\
C_{44} & i=l\neq k=j , \\
0 & \text{otherwise},
\end{array} \right. 
\label{eq:elasticity}
\end{equation}
and obeys the following symmetries
\begin{equation}
C_{ij}{}^{kl} = C_{ji}{}^{kl} = C_{ij}{}^{lk} = C_{kl}{}^{ij}\, .
\end{equation}
\end{subequations}
To obtain the relation between $\varepsilon_{\perp}$ and $\varepsilon_{\parallel}$ in terms of the elastic constants, the stiffness components entering Eq.~(\ref{eq:stress}) must be expressed in the standard cubic basis. The transformation law reads
\begin{equation}
C_{ij}^{\prime}{}^{kl} = \Lambda^{m}{}_{i} \Lambda^{n}{}_{j} C_{mn}{}^{rs}\Lambda_{r}{}^{k}\Lambda_{s}{}^{l}\, ,
\label{eq:LambdaTransf}
\end{equation}
with the rotation matrix
\begin{equation}
\Lambda = \left[
\begin{array}{ccc}
 1/\sqrt{6} & -1/\sqrt{2} & 1/\sqrt{3} \\
 1/\sqrt{6} & 1/\sqrt{2} & 1/\sqrt{3} \\
 -\sqrt{2/3} & 0 & 1/\sqrt{3} \\
\end{array}
\right] \in \mathrm{SO}(3)
\end{equation}
effecting the orthogonal rotation from the standard basis $\{ e_n \}$ to the quantum well basis $\{ e^{\prime}_n \}$, i.e., $e^{\prime}_n = \Lambda_n{}^m e_m$ (note $\Lambda_{n}{}^{m} = (\Lambda^{\mathrm{T}})^{m}{}_{n}$). Evaluating the transformation gives 
\begin{subequations}
\begin{equation}
C^{\prime}_{33}{}^{11} = C^{\prime}_{33}{}^{22} = \frac{1}{3}(C_{11} + 2C_{12} - 2C_{44})\, ,
\end{equation}
\begin{equation}
C^{\prime}_{33}{}^{33} = \frac{1}{3}(C_{11} + 2C_{12} + 4C_{44})\, ,
\end{equation}
\end{subequations}
which, when substituted into Eq.~(\ref{eq:stress}), yields the expression for $\varepsilon_{\perp}$ consistent with Eq.~(\ref{eq:eps_per}). The strain tensor represented in the standard basis (the form employed in Eq.~(\ref{eq:shift})) is calculated via the similarity transformation
\begin{equation}
\Lambda \varepsilon^{\prime} \Lambda^{\mathrm{T}} = \frac{1}{3}\left[
\begin{array}{ccc}
 2\varepsilon_{\parallel} + \varepsilon_{\perp} & \varepsilon_{\perp}-\varepsilon_{\parallel} & \varepsilon_{\perp}-\varepsilon_{\parallel} \\
 \varepsilon_{\perp}-\varepsilon_{\parallel} &2\varepsilon_{\parallel} + \varepsilon_{\perp} & \varepsilon_{\perp}-\varepsilon_{\parallel} \\
 \varepsilon_{\perp}-\varepsilon_{\parallel} & \varepsilon_{\perp}-\varepsilon_{\parallel} & 2\varepsilon_{\parallel} + \varepsilon_{\perp} \\
\end{array}
\right]. 
\end{equation}
To determine the energy $\delta E_{m}^{b}$ one must specify the components of the tensor $D_{m}^{b}$. These may be parametrized by the deformation potentials $\Xi_{d}^{b}$ and $\Xi_{u}^{b}$ as
 \begin{equation}
(D_{m}^{b})^{ik} = \Xi_{d}^{b}\delta^{ik} + \Xi_{u}^{b}\zeta^i\zeta^{k}\, ,
\label{eq:dxi}
\end{equation}
where $\bm{\mathit{\zeta}}$ is the unit vector that defines the valley $m \in \{ \Gamma, \mathrm{M} \}$ (cf. Fig.~\ref{subfig:IV-VIs_b}) and $\delta^{ik}$ is the Kronecker delta. Inserting Eq.~(\ref{eq:dxi}) into Eq.~(\ref{eq:shift}) results in
\begin{equation}
\delta E_{m}^{b} = \Xi_{d}^{b}\mathrm{Tr}(\varepsilon) + \Xi_{u}^{b}\bm{\mathit{\zeta}}^{\mathrm{T}}\varepsilon \bm{\mathit{\zeta}}\, .
\label{eq:trace}
\end{equation}
For the longitudinal $(\Gamma)$ valley $\bm{\mathit{\zeta}} = [111]\, /\sqrt{3}$, and for the oblique $(\mathrm{M})$ valley e.g. $\bm{\mathit{\zeta}} = [\bar{1}11]\, /\sqrt{3}$ (or any equivalent $\langle\bar{1}11\rangle\, /\sqrt{3}$ direction). Thus
\begin{subequations}
\begin{equation}
\delta E_{\Gamma}^{b} = \Xi_{d}^{b}\left(2\varepsilon_{\parallel} + \varepsilon_{\perp}\right) + \Xi_{u}^{b}\varepsilon_{\perp}\, ,
\end{equation}
\begin{equation}
\delta E^{b}_{\mathrm{M}} = \Xi_{d}^{b}\left(2\varepsilon_{\parallel} + \varepsilon_{\perp}\right) + \frac{\Xi_{u}^{b}}{9}\left(8\varepsilon_{\parallel} + \varepsilon_{\perp}\right) .
\end{equation}
\end{subequations}
The first term in Eq.~(\ref{eq:trace}) (valley-independent) describes the band-energy shift caused by the overall deformation of the crystal along the lattice-spanning directions. The second term discriminates between the $\Gamma$ and $\mathrm{M}$ valleys; it constitutes the directional (uniaxial) contribution and is proportional to the effective strain projected onto the unit vector $\bm{\mathit{\zeta}}$.

\begin{table*}[t]
\caption{\label{tab:strain_params} Experimental deformation potentials (in eV) and elastic constants (in $10^{10}\, \mathrm{N}/\mathrm{m}^2$) for PbSe \cite{Zasavitskii:2004_PRB} employed in the calculations. The parameters $\Xi_{d,u}^{v,c}$ were measured for PbTe and slightly adjusted to better reproduce existing experimental data for PbSe \cite{Kazakov:2025_PRB}. The isotropic parameter $\Xi_{\text{iso}}$ was measured directly for PbSe but could be applied only to the $[001]$~quantum~well.}
\begin{ruledtabular}
\begin{tabular}{cccccccc}
		$\Xi_{d}^{v}$ & $\Xi_{u}^{v}$ & $\Xi_{d}^{c}$ & $\Xi_{u}^{c}$ & $\Xi_{\text{iso}}$ & $C_{11}$ & $C_{12}$ & $C_{44}$ \Bstrut  \\ \hline
		\Tstrut $-8.93$ & $9.8$ & $-4.36$ & $7.9$ & $15.7$ & $14.18$ & $1.94$ & $1.749$ \\
	\end{tabular}
\end{ruledtabular}
\end{table*}

Similarly, the strain-induced band shifts may be obtained for the remaining growth directions since the crystal and deformation symmetries are encoded in the strain tensor. For quantum wells with growth axis $\hat{\bm{\mathit{e}}}_z \parallel [110]$ the transformation to the geometry-adapted basis $\left\{ [100], [1\bar{1}0], [110] \right\}$ and the resulting strain tensor are
\begin{subequations}
\begin{equation}
\Lambda = \left[
\begin{array}{ccc}
 0 & 1/\sqrt{2} & 1/\sqrt{2} \\
 0 & -1/\sqrt{2} & 1/\sqrt{2} \\
 1 & 0 & 0 \\
\end{array}
\right] ,
\end{equation}
\begin{equation}
\varepsilon = \Lambda \varepsilon^{\prime} \Lambda^{\mathrm{T}} = \frac{1}{2}\left[
\begin{array}{ccc}
 \varepsilon_{\parallel} + \varepsilon_{\perp} & \varepsilon_{\perp}-\varepsilon_{\parallel} & 0 \\
 \varepsilon_{\perp}-\varepsilon_{\parallel} &\varepsilon_{\parallel} + \varepsilon_{\perp} & 0 \\
 0 & 0 & 2\varepsilon_{\parallel} \\
\end{array}
\right].
\end{equation}
\end{subequations}
Evaluation of the transformed stiffness tensor components (using Eq.~(\ref{eq:LambdaTransf})) gives
\begin{subequations}
\begin{equation}
C^{\prime}_{33}{}^{11} = C^{\prime}_{33}{}^{22} = C_{12}\, ,
\end{equation}
\begin{equation}
C^{\prime}_{33}{}^{33} = \frac{1}{2}(C_{11} + C_{12} + 2C_{44})\, ,
\end{equation}
\end{subequations}
and the out-of-plane strain is related to the in-plane strain by
\begin{equation}
\varepsilon_{\perp} = -\frac{4C_{12}}{C_{11} + C_{12} + 2C_{44}}\varepsilon_{\parallel}\, .
\end{equation}
Under projection onto the $[110]$ direction the set of $\mathrm{L}$ valleys split into two subsets, $\left\{[111], [11\bar{1}]\right\}$ and $\left\{[\bar{1}11], [1\bar{1}1]\right\}$, corresponding to the $\mathrm{Y}$ and $\mathrm{M}$ points in Fig.~\ref{subfig:IV-VIs_b}, respectively. For these valleys, Eq.~(\ref{eq:trace}) yields
\begin{subequations}
\begin{equation}
\delta E_{\mathrm{Y}}^{b} = \Xi_{d}^{b}\left(2\varepsilon_{\parallel} + \varepsilon_{\perp}\right) + \Xi_{u}^{b}\varepsilon_{\parallel}\, ,
\end{equation}
\begin{equation}
\delta E^{b}_{\mathrm{M}} = \Xi_{d}^{b}\left(2\varepsilon_{\parallel} + \varepsilon_{\perp}\right) + \frac{\Xi_{u}^{b}}{3}\left(\varepsilon_{\parallel} + 2\varepsilon_{\perp}\right) .
\end{equation}
\end{subequations}
For growth with $\hat{\bm{\mathit{e}}}_z \parallel [001]$, and owing to the availability of experimental parameters, it is convenient to derive the strain-induced band gap shift  $\delta E_{\text{g}} = \delta E_{\mathrm{X}}^{c} - \delta E_{\mathrm{X}}^{v}$ for the four symmetry-equivalent projections of the $\mathrm{L}$ valleys onto the $\mathrm{X}$ points (cf. Fig.~\ref{subfig:IV-VIs_b}). In this orientation the quantum well and standard bases coincide so that $\varepsilon = \varepsilon^{\prime}$. Taking the valley direction $\bm{\mathit{\zeta}} = \langle 111\rangle\, /\sqrt{3}$ and evaluating Eq.~(\ref{eq:trace}) one finds
\begin{align}
& \delta E_{\text{g}} = \Xi_{d}\mathrm{Tr}(\varepsilon) + \Xi_{u}\bm{\mathit{\zeta}}^{\mathrm{T}}\varepsilon\bm{\mathit{\zeta}} = \Xi_{d}\left(2\varepsilon_{\parallel} + \varepsilon_{\perp}\right) \nonumber \\
& + \frac{\Xi_{u}}{3}\left(2\varepsilon_{\parallel} + \varepsilon_{\perp}\right) = \frac{1}{3}\left(3\Xi_{d} + \Xi_{u}\right)\left(2\varepsilon_{\parallel} + \varepsilon_{\perp}\right) ,
\end{align}
where $\Xi_{d,u} = \Xi_{d,u}^{c} - \Xi_{d,u}^{v}$ are the differences between the conduction- $(b=c)$ and valence- $(b=v)$ band deformation potentials. Introducing $\Xi_{\text{iso}} = 3\Xi_{d} + \Xi_{u}$ and using Eq.~(\ref{eq:stress}) leads to
\begin{equation}
\delta E_{\text{g}} = \frac{2}{3}\Xi_{\text{iso}}\frac{C_{11} - C_{12}}{C_{11}}\varepsilon_{\parallel}\, .
\end{equation}
Although experimental values of $\Xi_{d}^{b}$, $\Xi_{u}^{b}$, $\Xi_{\text{iso}}$, $C_{11}$, $C_{12}$ and $C_{44}$ are reported for many materials, such data are comparatively rare for mixed crystals. Therefore, the present calculations adopt the parameters measured for PbSe (Table~\ref{tab:strain_params}), neglecting modifications resulting from partial substitution of Pb by Sn.

Determination of the strain-tensor components (Eqs.~(\ref{eq:epsy})) requires the lattice constants of the quantum well and barrier materials, $a_{\text{QW}}(x)$ and $a_{\text{BR}}(y)$, as functions of the tin composition $x$ and the EuS fraction $y$. A commonly used approximation (and experimentally validated) is to assume linear composition dependence of the lattice constants (Vegard’s law):
\begin{subequations}
\label{eq:lattice}
\begin{align}
a_{\text{QW}}(x) & = a_{\text{PbSe}} - 0.1246 \cdot x \nonumber \\ & = 6.124 - 0.1246 \cdot x\ (\text{\AA})\, ,
\label{eq:lattice_QW}
\end{align}
\begin{align}
a_{\text{BR}}(y) & = a_{\text{PbSe}}\cdot(1-y) + a_{\text{EuS}} \cdot y \nonumber \\ & = 6.124\cdot(1-y) + 5.968 \cdot y\ (\text{\AA})\, .
\label{eq:lattice_BR}
\end{align}
\end{subequations}
The expression $a_{\text{QW}}(x)$ was derived in Ref.~\cite{Kritzman:2018_PRB} by linear regression of X-ray diffraction data for compositions up to $x \leqslant 0.25$. The lattice parameters of the pure constituents, $a_{\text{PbSe}}$ and $a_{\text{EuS}}$ (both rock-salt), are adopted from Refs.~\cite{Nimtz:1983} and \cite{WACHTER:1979_Alloys_and_Intermetallics}, respectively. It should be noted that the absence in nature of a stable rock-salt phase of SnSe (the thermodynamically stable phase of SnSe is orthorhombic) imposes a formal restriction on the range of validity of the fitted relation for $a_{\text{QW}}(x)$. However, given the capability of molecular-beam epitaxy (MBE) to stabilize lattice symmetry through suitable substrate and barrier selection, the linear Vegard-type interpolation of Eqs.~(\ref{eq:lattice}) is employed here beyond the fitted range ($x > 0.25$). A reasonable upper bound for the tin content is taken here as $x = 0.5$.

\begin{widetext}
With explicit forms of $\delta E_{m}^{b}$ and $\delta E_{\text{g}}$, the confinement potentials $V_{c,v}(z)$ for wells grown along $[111]$ and $[110]$ may be written as 
\begin{equation}
V_{c,v}(z) = \left\{ \begin{array}{ccc}
\pm E_{\text{g}}^{\text{QW}} /\, 2 + \delta E^{c,v}_m & \text{for} & |z|\leqslant d_{\text{QW}}\, /\, 2 , \\
\pm E_{\text{g}}^{\text{BR}} /\, 2 & \text{for} & |z| > d_{\text{QW}}\, /\, 2 , \\
\end{array} \right. 
\end{equation}
and for the $[001]$ orientation
\begin{equation}
V_{c,v}(z) = \left\{ \begin{array}{ccc}
\pm \left(E_{\text{g}}^{\text{QW}} + \delta E_{\text{g}}\right) /\, 2 & \text{for} & |z|\leqslant d_{\text{QW}}\, /\, 2 , \\
\pm E_{\text{g}}^{\text{BR}} /\, 2 & \text{for} & |z| > d_{\text{QW}}\, /\, 2 . \\
\end{array} \right. 
\end{equation}
The band gap $E_{\text{g}}$ is temperature- and composition-dependent. For Pb$_{1-x}$Sn$_{x}$Se the empirical relation from Ref.~\cite{Nimtz:1983} is used:
\begin{equation}
E_{\text{g}}^{\text{QW}}(T, x) = 125 - 930\cdot x + \sqrt{480 + 0.256\cdot (T[\mathrm{K}])^2}\ [\mathrm{meV}]\, ,
\label{eq:EgQW}
\end{equation}
whereas for (PbSe)$_{1-y}$(EuS)$_{y}$ the parametrization of Refs.~\cite{Maurice:1998_Phys_Stat_Sol,Simma:2012_APL} is adopted:
\begin{equation}
E_{\text{g}}^{\text{BR}}(T, y) = 146 + \frac{1 - 3\cdot y}{T[\mathrm{K}] + 40.7}\cdot 0.475\cdot(T[\mathrm{K}])^2 + 3000\cdot y\ [\mathrm{meV}]\, .
\label{eq:EgBR}
\end{equation}
The above formulae are formally applicable only in the regime of small values of the composition parameters $x$ and $y$. In particular, Eq.~(\ref{eq:EgQW}) was established for $x\leqslant 0.2$, while Eq.~(\ref{eq:EgBR}) is valid for $y < 0.14$ and was derived for the compound Pb$_{1-y}$Eu$_y$Se. Owing to the absence of alternative experimental parametrizations in the literature, the effect of sulfur incorporation on the barrier band gap $E_{\text{g}}^{\text{BR}}$ is neglected. This approximation is expected to be reasonable, as the band structures of EuS and EuSe are similar at the $\mathrm{L}$ point \cite{Schlipf:2013_PRB}. The relations (\ref{eq:EgQW}) and (\ref{eq:EgBR}) are therefore extrapolated beyond their original ranges of validity, into regions of $x$ and $y$ for which their quantitative accuracy may be diminished.

\end{widetext}


%


\end{document}